\documentclass[aps,amsmath,amssymb,prd,showpacs,floatfix,preprint,superscriptaddress,nofootinbib,12pt]{article}
\usepackage{jheppub}
\usepackage{bm}
\usepackage{ulem}

\usepackage{axodraw4j}
\usepackage{pstricks}
\usepackage{color}

\allowdisplaybreaks[3]

\begin{document}

\title{CFT data in the Gross-Neveu model}

\author[]{Mikhail Goykhman,}
\author[]{Ritam Sinha}
\affiliation[]{The Racah Institute of Physics, The Hebrew University of Jerusalem, \\ Jerusalem 91904, Israel}
\emailAdd{michael.goykhman@mail.huji.ac.il}
\emailAdd{ritam.sinha@mail.huji.ac.il}

\abstract{
We calculate CFT data
for the Gross-Neveu model in $2<d<4$ dimensions
at the next-to-leading
order in the $1/N$ expansion.
In particular, we make use of the background field method
to derive various conformal triangles involving the composite operator $s^2$,
for the Hubbard-Stratonovich field $s$. We then apply these conformal triangles
to obtain the corresponding OPE coefficients.
}

\maketitle

\section{Introduction}
\label{sec:introduction}

Some of the most interesting phenomena in nature are associated with the
regime of the strong interaction and therefore are not accessible to the standard perturbative treatment. 
Particularly, strongly-coupled physics occurs
in the critical regime and is described by conformal field theories (CFTs) \cite{Polyakov:1970xd}. From the standpoint
of fundamental physics, such a regime is reached at the end of the renormalization group (RG)
flow of a quantum field theory (QFT), either in the deep infra-red (IR) or in the asymptotically ultra-violet (UV) regime.
Moreover, different systems
can have their RG flows terminate at the same
CFT, in a manifestation of the renowned phenomenon of critical universality.

The universality principle served as one of the inspirations
to classify CFTs based on the algebra of primary operators and their observable properties,
such as scaling dimensions 
and operator product expansion (OPE) coefficients, without necessarily specifying
the Lagrangian of the underlying theory.
In particular, the microscopic specifics of the underlying theory are ignored
while the symmetries of the system and general consistency conditions
(such as unitarity)
constraining the available space of parameters come forward.
The corresponding program is known as the
conformal bootstrap, and it has been under active development over recent years
\cite{Parisi:1972zm,Polyakov:1974gs,Ferrara:1973yt,ElShowk:2012ht,Simmons-Duffin:2016gjk}.

This paper is motivated by the desire to expand our
understanding of critical dynamics in $d=3$ dimensions.
The well-known example of a three-dimensional critical system is furnished
by the IR fixed point of the $O(n)$ vector model.
This model can be viewed as a continuum limit description of the critical $n$-vector model on a lattice
\cite{Stanley:1968ef,Guida:1998bx,ElShowk:2012ht}.
The latter in turn generalizes the three-dimensional Ising model, describing the second-order
phase transition of a ferromagnet.
The fermionic counterpart of the $O(n)$ vector model is given by the $U(n)$
models with quartic fermion couplings, which we choose as the main focus of this paper.

The fixed point of the $O(n)$ vector model
exists in a perturbative Wilson-Fisher regime when the model is considered in $4-\epsilon$ dimensions
for small values of $\epsilon$.
The three-dimensional physics (after a proper re-summation)
is rather well approximated by setting $\epsilon = 1$ \cite{Wilson:1971dc}.
Similarly, the $U(n)$ fermionic model with the four-fermion Gross-Neveu interaction
is asymptotically free in two dimensions\footnote{In fact, it possesses a number of remarkable
properties in two dimensions, such as the dynamical breaking of chiral symmetry and generation of the IR
scale via the dimensional transmutation, making it a 2d toy-model of quantum
chromodynamics \cite{Gross:1974jv}.} but
possesses the Wilson-Fisher type of fixed point in the UV limit in $2+\epsilon$ dimensions
\cite{Gross:1974jv,ZinnJustin:1991yn}.

While the critical vector models are non-perturbative in general $d$,
a popular approach to study them is given by the $1/N$ expansion,
around an infinitely large number $N\rightarrow\infty$
of the degrees of freedom of the system (see \cite{Parisi:1975im} and references
therein).
The $1/N$ expansion and conformal bootstrap techniques are well suited
to study strongly-coupled critical regime and have proven to be remarkably successful 
methods for extracting  the CFT data such as the scaling dimensions and the OPE coefficients, see \cite{Vasiliev:1981yc,Vasiliev:1981dg,Vasiliev:1975mq,Vasiliev:1982dc,Gracey:1990wi,ZinnJustin:1991yn,Lang:1991kp,Gracey:1992cp,Vasiliev:1992wr,Vasiliev:1993pi,Gracey:1993kb,Gracey:1993kc,Lang:1993ct,Petkou:1995vu,Petkou:1994ad,Derkachov:1993uw,Derkachov:1997ch,Leonhardt:2003du,Fei:2014yja,Fei:2014xta,Gracey:2015tta,Manashov:2016uam,Diab:2016spb,Giombi:2016fct,Fei:2016sgs,Gracey:2016mio,Manashov:2017rrx,Gracey:2018ame,Alday:2019clp,Giombi:2019upv,Goykhman:2019kcj,Goykhman:2020ffn} for an incomplete list of related references.\footnote{
While the quartic coupling in the $O(n)$ vector model in $4<d<6$ dimensions, and the
Gross-Neveu coupling in $2<d<4$ dimensions are non-renormalizable by power counting, these
theories are renormalizable at each order in the $1/N$ expansion \cite{Parisi:1975im,Rosenstein:1988pt}. 
See also \cite{Goykhman:2019kcj} where consistency check for the existence of a fixed point in the $O(n)$ vector
model in $2<d<6$ dimensions was carried out by examining the Callan-Symanzik equations.}

Recently the power of the background field method was emphasized in the
context of large-$N$ vector models \cite{Goykhman:2020ffn}. It was shown that formally fixing some of the
degrees of freedom to their non-dynamical background values provides
a simple short-cut to the calculation of effective vertices at sub-leading orders in the $1/N$
expansion. 
In particular, this procedure allows one to easily extract finite parts of the
three-point correlation functions (OPE coefficients).
This is contrasted with the typical
focus of the research in large $N$ critical vector models on the calculation of the critical exponents.
In the literature, the background field method has been used effectively
for calculations in cases where the field acquires a vacuum expectation value (v.e.v.) as
a result of spontaneous symmetry breaking \cite{Goldstone:1962es}.
However, in the case of \cite{Goykhman:2020ffn}, as well as in this paper,
the considered fields do not necessarily acquire a v.e.v.,
and therefore the method should be viewed only as a calculation tool.

In this paper, we further establish the power of the background
field method for the study of critical vector models. To this end, we will scrutinize the 
UV critical regime of the $U(n)$ fermionic model with the Gross-Neveu 
interaction, and compute the full $s^2 s s$ and $s^2 \bar\psi\psi$ effective
vertices for the fermion $\psi$, the Hubbard-Stratonovich field $s$, and the
composite operator $s^2$ in terms
of the corresponding conformal triangles.\footnote{See \cite{Vasiliev:1993pi} for the earlier calculation of the $s^2 s s$
conformal triangle, which focused on its singular part for the purpose of calculating the anomalous
dimension $\gamma_{s^2}$. We extend their result by adding the finite part of this
conformal triangle, essential for the calculation of the OPE coefficients.} This is the first time that 
the background field method
is used to calculate correlation functions involving a composite operator.
In a simple way, this method provides a very powerful technique to calculate the finite part of non-planar diagrams contributing to these conformal triangles. We further use these conformal triangles
to calculate the OPE coefficients appearing in the correlation functions
involving the operator $s^2$, including the two-point function $\langle s^2 s^2\rangle$,
and the three-point functions $\langle s^2\bar\psi\psi\rangle$ and $\langle s^2ss\rangle$.

It is well known that the UV fixed point of the GN model in $2<d<4$ dimensions is described by the same
CFT as the IR fixed point of the Gross-Neveu-Yukawa (GNY) model \cite{ZinnJustin:1991yn}.
Such an example of critical universality can also be interpreted in terms of the GNY model
being a UV completion of the GN model. Moreover, the GNY model can be studied
perturbatively in $d=4-\epsilon$ dimensions, and the corresponding CFT data should
match the CFT data of the GN model \cite{ZinnJustin:1991yn}. We perform such a consistency
check on all of our results obtained in this paper.\footnote{In a similar spirit, the $O(N)$ vector model
in $4<d<6$ dimensions has been conjectured to have a UV completion in terms of the vector model coupled
to a dynamical scalar field with cubic interaction. This conjecture has received numerous
verifications up to the fourth order in perturbation theory \cite{Fei:2014yja,Fei:2014xta,Gracey:2015tta,Diab:2016spb,Giombi:2019upv,Goykhman:2019kcj}.}

 The rest of this paper is organized as follows. In section~\ref{sec:setup} we set up our conventions
 and review the known results in the literature that will be useful for the purposes of this paper.
 In section~\ref{sec:s^2 bar psi psi} we use the background field method to calculate
 the new conformal triangles up to the next-to-leading order in the $1/N$
 expansion. We derive the $s^2\bar\psi\psi$ conformal triangle in subsection~\ref{subsec:s2psipsi},
 and the $s^2 s s$ conformal triangle in subsection~\ref{subsec:s2ss}.
In section~\ref{sec:anomalous dim of s2} we carry out the calculation of the $\langle s^2 s s\rangle$ three-point
function at the next-to-leading order in the $1/N$ expansion. In the process,
we derive various OPE coefficients. In section~\ref{sec:s2psipsi} we calculate
the $\langle s^2\bar\psi\psi\rangle$ three-point function at the next-to-leading order in the $1/N$
expansion. In section~\ref{sec:s2} we demonstrate how the $s^2 s s$ conformal triangle
can be calculated from the $\langle s^2 s^2\rangle$ two-point function, which in particular
serves as a non-trivial consistency check for our results. We conclude with the discussion
in section~\ref{sec:discussion}, where we also outline possible future directions.

\section{Set-up}
\label{sec:setup}

Consider the $U(n)$-invariant fermionic model in $2\leq d\leq 4$ dimensions
with the quartic Gross-Neveu interaction,
\begin{equation}
\label{GN original action}
S = \int d^dx\,\left(\bar\psi\gamma^\mu\partial_\mu\psi
+\frac{g}{N}\,\left(\bar\psi\psi\right)^2\right)\,,
\end{equation}
where $\psi$ is the $n$-component multiplet of Dirac fermions.
According to the standard conventions
$N=n\,\textrm{tr}\mathbb{I}$, where $\mathbb{I}$
is the unit matrix in the $2^{[d/2]}$-dimensional space of Dirac spinors.
We will be working in the Euclidean signature, with Hermitian gamma-matrices, 
$(\gamma^\mu)^\dagger = \gamma^\mu$, such that
$\{\gamma^\mu,\gamma^\nu\}=2\delta^{\mu\nu}\,\mathbb{I}$.
Below in this paper we will skip keeping explicit track of the $U(n)$ indices
where it does not cause a confusion.\footnote{In particular, when writing down
the Feynman rules, we will omit
the Kronecker delta-symbols for the $U(n)$ indices.}

Using the standard trick, known as the Hubbard-Stratonovich transformation,
we can rewrite the action (\ref{GN original action}) in terms of the original fermions
$\psi$ as well as an auxiliary scalar field $s$,
\begin{equation}
\label{GN action}
S = \int d^dx\,\left(\bar\psi\gamma^\mu\partial_\mu\psi -
\frac{1}{4g}\,s^2+\frac{1}{\sqrt{N}}\,s\bar\psi\psi \right)\,.
\end{equation}
The Hubbard-Stratonovich field $s$ becomes dynamical due to the fermion
loop diagrams.
The model (\ref{GN action}) is believed to reach a non-trivial UV fixed point in $2<d<4$ dimensions \cite{ZinnJustin:1991yn},\footnote{Such a fixed point can be studied perturbatively in $2+\epsilon$ dimensions.}
and we will be studying the corresponding CFT.

We will use the following Feynman rules for the bare propagators of the Dirac fermion
and the Hubbard-Stratonovich field $s$, and the leading order interaction vertex:
\begin{center}
  \begin{picture}(300,67) (24,-12)
    \SetWidth{1.0}
    \SetColor{Black}
    \Line[arrow,arrowpos=0.5,arrowlength=5,arrowwidth=2,arrowinset=0.2](34,34)(98,34)
    \Vertex(34,34){2}
    \Vertex(98,34){2}
    \Text(106,25)[lb]{\scalebox{1}{$=\frac{C_\psi\,x^\mu \gamma_\mu}{|x|^d}$}}
    \Text(106,-15)[lb]{\scalebox{1}{$=\frac{C_s}{|x|^2}$}}
    \Text(98,26)[lb]{\scalebox{0.8}{$x$}}
    \Text(24,26)[lb]{\scalebox{0.8}{$0$}}
    \Line[](32,-6)(96,-6)
    \Text(21,-14)[lb]{\scalebox{0.8}{$0$}}
    \Text(98,-14)[lb]{\scalebox{0.8}{$x$}}
    \Vertex(32,-6){2}
    \Vertex(96,-6){2}
    \Line[arrow,arrowpos=0.5,arrowlength=5,arrowwidth=2,arrowinset=0.2](185,34)(230,14)
    \Line[arrow,arrowpos=0.5,arrowlength=5,arrowwidth=2,arrowinset=0.2](230,14)(185,-6)
    \Line[](230,14)(272,14)
    \Vertex(230,14){4}
    \Text(289,5)[lb]{\scalebox{1}{$=-\frac{1}{\sqrt{N}}$}}
  \end{picture}
\end{center}
A fermionic loop generates the factor of $-n\,\textrm{tr}(\mathbb{I})=-N$,
where the minus sign appears as a consequence of
 the Wick contractions due to the anti-commuting nature of the fermion field.
We will also use the following notation for the regularized Hubbard-Stratonovich propagator:
\begin{center}
  \begin{picture}(300,17) (24,-12)
    \SetWidth{1.0}
    \SetColor{Black}
    \Line[](32,-6)(96,-6)
    \Text(21,-14)[lb]{\scalebox{0.8}{$0$}}
    \Text(98,-14)[lb]{\scalebox{0.8}{$x$}}
    \Text(47,0)[lb]{\scalebox{0.8}{$2\Delta_s+\delta$}}
    \Vertex(32,-6){2}
    \Vertex(96,-6){2}
     \Text(106,-15)[lb]{\scalebox{1}{$=\frac{C_s\,\mu^{-\delta}}{|x|^{2\Delta_s+\delta}}$}}
  \end{picture}
\end{center}
where $\delta$ is the regularization parameter which is taken
to zero at the end of calculation \cite{Vasiliev:1975mq}, $\mu$ is the renormalization mass scale,
and $\Delta_s=1+\gamma_s$ is dimension of the Hubbard-Stratonovich field,
which acquires the anomalous term $\gamma_s$.
Our notations for the scalar and fermion lines
with a general exponent and a unit amplitude will be as follows:
\begin{center}
  \begin{picture}(300,30) (24,20)
    \SetWidth{1.0}
    \SetColor{Black}
    \Line[arrow,arrowpos=0.5,arrowlength=5,arrowwidth=2,arrowinset=0.2](34,34)(98,34)
    \Vertex(34,34){2}
    \Vertex(98,34){2}
    \Text(106,25)[lb]{\scalebox{1}{$=\frac{x^\mu \gamma_\mu}{|x|^{2a+1}}$}}
    \Text(266,25)[lb]{\scalebox{1}{$=\frac{1}{|x|^{2a}}$}}
    \Text(98,26)[lb]{\scalebox{0.8}{$x$}}
    \Text(24,26)[lb]{\scalebox{0.8}{$0$}}
    \Text(62,40)[lb]{\scalebox{0.8}{$2a$}}
    \Line[](192,34)(256,34)
    \Text(181,26)[lb]{\scalebox{0.8}{$0$}}
    \Text(222,40)[lb]{\scalebox{0.8}{$2a$}}
    \Text(258,26)[lb]{\scalebox{0.8}{$x$}}
    \Vertex(192,34){2}
    \Vertex(256,34){2}
  \end{picture}
\end{center}
According to this notation, we will typically skip explicitly labeling exponents of the
bare propagators of the fundamental fermion and the Hubbard-Stratonovich field.

The full fermion $\psi$ and Hubbard-Stratonovich $s$ propagators, including
the anomalous dimensions $\gamma_{\psi, s}$ as well as corrections to the amplitudes
$A_{\psi, s}$, will be denoted with solid
blobs, and are given by
\begin{center}
  \begin{picture}(258,35) (94,-9)
    \SetWidth{1.0}
    \SetColor{Black}
    \Line[arrow,arrowpos=0.2,arrowlength=5,arrowwidth=2,arrowinset=0.2](167,4)(272,4)
    \Line[arrow,arrowpos=0.8,arrowlength=5,arrowwidth=2,arrowinset=0.2](167,4)(272,4)
    \Vertex(167,4){2}
    \Vertex(272,4){2}
    \Text(160,-5)[lb]{\scalebox{0.8}{$0$}}
    \Text(277,-5)[lb]{\scalebox{0.8}{$x$}}
    \GOval(221,5)(13,13)(0){0.882}
    \Text(287,-5)[lb]{\scalebox{1}{$=C_\psi(1+A_\psi)\,\mu^{-2\gamma_\psi}\,
    \frac{x^\mu\gamma_\mu}{|x|^{d+2\gamma_\psi}}$}}
    \Text(85,-1)[lb]{\scalebox{1}{$\langle \psi(0)\bar\psi(x)\rangle = $}}
  \end{picture}
\end{center}
\begin{center}
  \begin{picture}(258,35) (94,-9)
    \SetWidth{1.0}
    \SetColor{Black}
    \Line[](167,4)(272,4)
    \Vertex(167,4){2}
    \Vertex(272,4){2}
    \GOval(221,5)(13,13)(0){0.882}
    \Text(160,-5)[lb]{\scalebox{0.8}{$0$}}
    \Text(277,-5)[lb]{\scalebox{0.8}{$x$}}
    \Text(287,-5)[lb]{\scalebox{1}{$=C_s(1+A_s)\,\mu^{-2\gamma_s}\,\frac{1}{|x|^{2+2\gamma_s}}$}}
    \Text(85,-1)[lb]{\scalebox{1}{$\langle s(x)s(0)\rangle = $}}
  \end{picture}
\end{center}
The full propagator of the composite field $s^2$ will be denoted as
\begin{center}
  \begin{picture}(258,35) (94,-9)
    \SetWidth{1.0}
    \SetColor{Black}
    \Line[](167,4)(272,4)
    \CBox(163,1)(169,7){Black}{Black}
    \CBox(268,1)(274,7){Black}{Black}
    \GOval(221,5)(13,13)(0){0.882}
    \Text(160,-8)[lb]{\scalebox{0.8}{$0$}}
    \Text(277,-8)[lb]{\scalebox{0.8}{$x$}}
    \Text(287,-5)[lb]{\scalebox{1}{$=C_{s^2}(1+A_{s^2})\,\mu^{-2\gamma_{s^2}}\,\frac{1}{|x|^{4+2\gamma_{s^2}}}$}}
    \Text(85,-1)[lb]{\scalebox{1}{$\langle s(x)^2s(0)^2\rangle = $}}
  \end{picture}
\end{center}
where we defined
\begin{equation}
\label{Cs2 def}
C_{s^2} = 2C_s^2
\end{equation}
to be the leading order amplitude of the $\langle s^2s^2\rangle$ propagator.
The leading order propagator amplitudes are given by \cite{ZinnJustin:1991yn}
\begin{align}
C_\psi=\frac{\Gamma\left(\frac{d}{2}\right)}{2\pi^\frac{d}{2}}\,,\qquad
C_s = -\frac{2^d\sin\left(\frac{\pi d}{2}\right)\Gamma\left(\frac{d-1}{2}\right)}{\pi^\frac{3}{2}\,
\Gamma\left(\frac{d}{2}-1\right)}\,,
\end{align}
The anomalous dimensions at the next-to-leading order are \cite{ZinnJustin:1991yn}
\begin{equation}
\gamma_\psi = -\frac{1}{N}\,\frac{2^{d-1}\sin\left(\frac{\pi d}{2}\right)\Gamma\left(\frac{d-1}{2}\right)}
{\pi^\frac{3}{2}\,d\,\Gamma\left(\frac{d}{2}-1\right)}+{\cal O}\left(1/N^2\right)\,,\qquad
\gamma_s = -4\frac{d-1}{d-2}\,\gamma_\psi+{\cal O}\left(1/N^2\right)\,,\label{gamma psi and s}
\end{equation}
and $1/N$ corrections to the propagators amplitudes are \cite{Manashov:2017rrx}
\begin{equation}
A_\psi = -\frac{2}{d}\,\gamma_\psi+{\cal O}\left(1/N^2\right)
\,,\quad A_s = -\left(H_{d-2}+\frac{2}{d}+\pi\,\cot\left(\frac{\pi d}{2}\right)\right)\,\gamma_s+{\cal O}\left(1/N^2\right)\,,
\end{equation}
where $H_n$ is $n$th harmonic number.

We will also need the $s\bar\psi\psi$ effective vertex, represented by the corresponding conformal triangle \cite{Polyakov:1970xd},
and denoted with a solid blob
\begin{center}
  \begin{picture}(400,88) (19,-10)
    \SetWidth{1.0}
    \SetColor{Black}
    \GOval(23,24)(14,15)(0){0.882}
    \Line[](23,65)(23,38)
    \Line[arrow,arrowpos=0.5,arrowlength=5,arrowwidth=2,arrowinset=0.2](14,12)(-2,-7)
    \Line[arrow,arrowpos=0.5,arrowlength=5,arrowwidth=2,arrowinset=0.2](48,-7)(32,12)
    \Line[](172,76)(172,38)
    \Line[arrow,arrowpos=0.5,arrowlength=5,arrowwidth=2,arrowinset=0.2](171,38)(144,3)
    \Line[arrow,arrowpos=0.5,arrowlength=5,arrowwidth=2,arrowinset=0.2](144,3)(115,-9)
    \Line[arrow,arrowpos=0.5,arrowlength=5,arrowwidth=2,arrowinset=0.2](227,-9)(195,3)
    \Vertex(145,2){4}
    \Line[arrow,arrowpos=0.5,arrowlength=5,arrowwidth=2,arrowinset=0.2](195,3)(173,38)
    \Line[](198,3)(144,3)
    \Vertex(194,4){4}
    \Vertex(172,39){4}
    \Text(60,20)[lb]{\scalebox{0.8}{$=-\frac{Z_{s\bar\psi\psi}}{\sqrt{N}}\,\mu^{2\gamma_\psi+\gamma_s}\times$}}
    \Text(143,20)[lb]{\scalebox{0.8}{$2\alpha$}}
    \Text(190,20)[lb]{\scalebox{0.8}{$2\alpha$}}
    \Text(165,-10)[lb]{\scalebox{0.8}{$2\beta$}}
    \Text(178,43)[lb]{\scalebox{0.8}{$s(x_1)$}}
    \Text(133,-16)[lb]{\scalebox{0.8}{$\bar\psi(x_2)$}}
    \Text(190,-16)[lb]{\scalebox{0.8}{$\psi(x_3)$}}
    \Text(219,15)[lb]{\scalebox{1}{$=-\frac{Z_{s\bar\psi\psi}}{\sqrt{N}}\int d^dx_{1,2,3}
    \frac{ x_{21}^\mu\gamma_\mu
    x_{13}^\nu\gamma_\nu \mu^{2\gamma _ \psi  + \gamma _ s}s(x_1)\bar\psi(x_2)\psi(x_3)}{(|x_{12}||x_{13}|)^{2\alpha+1}|x_{12}|^{2\beta}}$}}
  \end{picture}
\end{center}
where the exponents
\begin{equation}
\alpha = \frac{d-1-\gamma_s}{2}\,,\qquad
\beta = 1 -\gamma_\psi +\frac{\gamma_s}{2}
\end{equation}
are such that the integration vertices $x_{1,2,3}$
become unique when the propagators are attached to them.
The $\bar\psi\psi s$ conformal triangle diagram is a sum of the leading order tree-level diagram, the sub-leading
loop corrections, and the $\bar\psi\psi s$ vertex counter-term.
In particular, the $\bar\psi\psi s$ vertex counter-term cancels the divergences from
the sub-leading order loop diagram,
see \cite{Goykhman:2020ffn} for the detailed explanation.

To the next-to-leading order in $1/N$ we can expand
\begin{equation}
Z_{s\bar\psi\psi} = Z_{s\bar\psi\psi}^{(0)}\left(1+\delta Z_{s\bar\psi\psi}+{\cal O}(1/N)\right)\,,
\end{equation}
where \cite{Manashov:2017rrx}\footnote{See also \cite{Goykhman:2020ffn} for a recent derivation.}
\begin{align}
Z_{s\bar\psi\psi}^{(0)} &=-\frac{(d-2)\Gamma\left(\frac{d}{2}\right)^2}{4\pi^d}
\,(2\gamma_\psi+\gamma_s)\,,\\
\delta Z_{s\bar\psi\psi} &=-\frac{2}{d-2}\,\gamma_s\,.
\end{align}

For the analysis of loop diagrams we
we will be using the following propagator merging relations \cite{Gracey:1990wi}
\begin{center}
  \begin{picture}(332,105) (13,-40)
    \SetWidth{1.0}
    \SetColor{Black}
    \Line[arrow,arrowpos=0.5,arrowlength=5,arrowwidth=2,arrowinset=0.2](18,48)(67,48)
    \Line[](66,48)(115,48)
    \Vertex(66,48){4}
    \Line[arrow,arrowpos=0.5,arrowlength=5,arrowwidth=2,arrowinset=0.2](146,48)(195,48)
    \Line[arrow,arrowpos=0.5,arrowlength=5,arrowwidth=2,arrowinset=0.2](18,11)(67,11)
    \Vertex(66,11){4}
    \Line[arrow,arrowpos=0.5,arrowlength=5,arrowwidth=2,arrowinset=0.2](67,11)(116,11)
    \Line[](144,11)(193,11)
    \Text(38,53)[lb]{\scalebox{0.8}{$2\Delta_1$}}
    \Text(86,53)[lb]{\scalebox{0.8}{$2\Delta_2$}}
    \Text(38,16)[lb]{\scalebox{0.8}{$2\Delta_1$}}
    \Text(86,16)[lb]{\scalebox{0.8}{$2\Delta_2$}}
    \Text(140,53)[lb]{\scalebox{0.8}{$2(\Delta_1+\Delta_2)-d$}}
    \Text(138,16)[lb]{\scalebox{0.8}{$2(\Delta_1+\Delta_2)-d$}}
    \Text(210,45)[lb]{\scalebox{0.8}{$\times\pi^\frac{d}{2}\,A(\Delta_2)\,V(\Delta_1,d-\Delta_1-\Delta_2)$}}
    \Text(210,10)[lb]{\scalebox{0.8}{$\times (-\pi^\frac{d}{2})\,A(d-\Delta_1-\Delta_2)\,V(\Delta_1,\Delta_2)\times\mathbb{I}$}}
    \Text(127,45)[lb]{\scalebox{0.8}{$=$}}
    \Text(127,10)[lb]{\scalebox{0.8}{$=$}}
    \Vertex(18,48){2}
    \Vertex(18,11){2}
    \Vertex(115,48){2}
    \Vertex(116,11){2}
    \Vertex(146,48){2}
    \Vertex(195,48){2}
    \Vertex(146,11){2}
    \Vertex(195,11){2}
    \Text(38,-21)[lb]{\scalebox{0.8}{$2\Delta_1$}}
    \Text(86,-21)[lb]{\scalebox{0.8}{$2\Delta_2$}}
    \Text(138,-21)[lb]{\scalebox{0.8}{$2(\Delta_1+\Delta_2)-d$}}
    \Text(127,-28)[lb]{\scalebox{0.8}{$=$}}
    \Text(210,-29)[lb]{\scalebox{0.8}{$\times U(\Delta_1,\Delta_2,d-\Delta_1-\Delta_2)$}}
    \Vertex(18,-26){2}
    \Vertex(116,-26){2}
    \Vertex(146,-26){2}
    \Vertex(195,-26){2}
    \Line[](18,-26)(67,-26)
    \Line[](67,-26)(116,-26)
    \Line[](144,-26)(193,-26)
    \Vertex(66,-26){4}
  \end{picture}
\end{center}
where we defined
\begin{align}
A(\Delta) &= \frac{\Gamma\left(\frac{d}{2}-\Delta\right)}{\Gamma(\Delta)}\,,\qquad
V(\Delta_1,\Delta_2) = \frac{\Gamma\left(\frac{d+1}{2}-\Delta_1\right)}{\Gamma(\Delta_1+\frac{1}{2})}
\frac{\Gamma\left(\frac{d+1}{2}-\Delta_2\right)}{\Gamma(\Delta_2+\frac{1}{2})}\,,\\
U(\Delta_1,\Delta_2,\Delta_2) &= \pi^\frac{d}{2}\,A(\Delta_1)A(\Delta_2)A(\Delta_3)\,,
\end{align}
and the uniqueness relations for $\Delta_1+\Delta_2+\Delta_3=d$ \cite{DEramo:1971hnd,Symanzik:1972wj,Gracey:1990wi},
\begin{align}
\label{scalar uniqueness}
\int d^d x_4\,\frac{1}
{|x_{14}|^{2\Delta_1}|x_{24}|^{2\Delta_2}|x_{34}|^{2\Delta_3}}&=
\frac{U(\Delta_1,\Delta_2,\Delta_3)}
{|x_{12}|^{d-2\Delta_3}|x_{13}|^{d-2\Delta_2}|x_{23}|^{d-2\Delta_1}}\,,\\
\label{fermion uniqueness}
\int d^d x_4\,\frac{x_{24}^\mu \gamma_\mu x_{41}^\nu \gamma_\nu}
{|x_{14}|^{2\Delta_1+1}|x_{24}|^{2\Delta_2+1}|x_{34}|^{2\Delta_3}}&=
\frac{\pi^\frac{d}{2}A(\Delta_3)V(\Delta_1,\Delta_2)x_{31}^\mu\gamma_\mu x_{23}^\nu\gamma_\nu}
{|x_{12}|^{d-2\Delta_3}|x_{13}|^{d-2\Delta_2+1}|x_{23}|^{d-2\Delta_1+1}}\,.
\end{align}
Loops in position space are simply additive:\footnote{Notice that the minus sign
in the r.h.s. of the last loop is independent of the minus
sign in the factor of $-N$ appearing in the Feynman rule for the fermionic loop.}
\begin{center}
  \begin{picture}(493,22) (30,4)
    \SetWidth{1.0}
    \SetColor{Black}
    \Arc[clock](64.5,-28.8)(57.802,124.212,55.788)
    \Arc[](65,78.938)(67.945,-118.097,-61.903)
    \Line[](113,19)(161,19)
    \Arc[arrow,arrowpos=0.5,arrowlength=5,arrowwidth=2,arrowinset=0.2,clock](224.5,-28.8)(57.802,124.212,55.788)
    \Arc[](225,78.938)(67.945,-118.097,-61.903)
    \Line[arrow,arrowpos=0.5,arrowlength=5,arrowwidth=2,arrowinset=0.2](273,19)(321,19)
    \Arc[arrow,arrowpos=0.5,arrowlength=5,arrowwidth=2,arrowinset=0.2,clock](385.5,-27.8)(57.802,124.212,55.788)
    \Arc[arrow,arrowpos=0.5,arrowlength=5,arrowwidth=2,arrowinset=0.2,clock](385,79)(68,-61.928,-118.072)
    \Line[](433,19)(481,19)
    \Vertex(33,19){2}
    \Vertex(96,19){2}
    \Vertex(113,19){2}
    \Vertex(161,19){2}
    \Vertex(193,19){2}
    \Vertex(256,19){2}
    \Vertex(353,19){2}
    \Vertex(418,19){2}
    \Vertex(273,19){2}
    \Vertex(321,19){2}
    \Vertex(433,19){2}
    \Vertex(481,19){2}
    \Text(488,15)[lb]{\scalebox{0.8}{$\times (-\mathbb{I})$}}
    \Text(423,18)[lb]{\scalebox{0.8}{$=$}}
    \Text(262,18)[lb]{\scalebox{0.8}{$=$}}
    \Text(101,18)[lb]{\scalebox{0.8}{$=$}}
    \Text(60,32)[lb]{\scalebox{0.8}{$2\Delta_1$}}
    \Text(60,-1)[lb]{\scalebox{0.8}{$2\Delta_2$}}
    \Text(115,23)[lb]{\scalebox{0.8}{$2(\Delta_1+\Delta_2)$}}
    \Text(220,34)[lb]{\scalebox{0.8}{$2\Delta_1$}}
    \Text(222,-1)[lb]{\scalebox{0.8}{$2\Delta_2$}}
    \Text(273,23)[lb]{\scalebox{0.8}{$2(\Delta_1+\Delta_2)$}}
    \Text(379,34)[lb]{\scalebox{0.8}{$2\Delta_1$}}
    \Text(379,-1)[lb]{\scalebox{0.8}{$2\Delta_2$}}
    \Text(433,23)[lb]{\scalebox{0.8}{$2(\Delta_1+\Delta_2)$}}
  \end{picture}
\end{center}

\section{Conformal triangles}
\label{sec:s^2 bar psi psi}

The structure of three-point functions in CFTs is completely fixed
by the conformal symmetry.
It is usually convenient to decompose contributions to the three-point functions
(and the corresponding OPE coefficients)
into the terms originating from the corresponding conformal
triangle \cite{Polyakov:1970xd}, and the terms due to the amplitudes of the propagators 
attached to the conformal triangle  \cite{Goykhman:2019kcj}.
In this section we will make use of the background field method to calculate
the $s^2\bar\psi \psi$ and $s^2ss$ conformal triangles.

The results obtained in this section will be used below
in section~\ref{sec:anomalous dim of s2} to derive the $1/N$ correction $A_{s^2}$
to the amplitude of the $\langle s^2 s^2\rangle$ propagator,
and the $1/N$ corrections to the three-point function $\langle s^2 ss\rangle$.
Finally, in section~\ref{sec:s2psipsi}
we will use these results to obtain
the next-to-leading order corrections to the three-point
function $\langle s^2\bar\psi\psi\rangle$.

\subsection{$s^2 \bar\psi\psi$ conformal triangle}
\label{subsec:s2psipsi}

The $s^2\bar\psi \psi$ conformal triangle
can be represented diagrammatically as
\begin{center}
  \begin{picture}(102,108) (11,10)
    \SetWidth{1.0}
    \SetColor{Black}
    \Line[](72,96)(32,32)
    \Line[](112,32)(72,96)
    \Line[arrow,arrowpos=0.5,arrowlength=5,arrowwidth=2,arrowinset=0.2](112,32)(32,32)
    \Vertex(32,32){4}
    \Vertex(112,32){4}
    \Vertex(72,96){4}
    \Text(70,20)[lb]{\scalebox{0.8}{$2b$}}
    \Text(32,64)[lb]{\scalebox{0.8}{$2a$}}
    \Text(104,64)[lb]{\scalebox{0.8}{$2a$}}
    \Text(64,105)[lb]{\scalebox{0.8}{$s(x_1')^2$}}
    \Text(14,15)[lb]{\scalebox{0.8}{$\bar\psi(x_2')$}}
    \Text(115,15)[lb]{\scalebox{0.8}{$\psi(x_3')$}}
    \Text(-100,60)[lb]{\scalebox{1}{$-\frac{1}{N}\,Z_{s^2\bar\psi\psi}\,\mu^{2\gamma_\psi+\gamma_{s^2}}\quad\times$}}
  \end{picture}
\end{center}
Here we have introduced
\begin{equation}
\label{a and b for conformal triangle}
a = \frac{d-\gamma_{s^2}}{2}-1\,,\qquad b=\frac{3+\gamma_{s^2}}{2}-\gamma_\psi\,.
\end{equation}
The structure of the conformal triangle is such that when the full propagators of the composite
operator $s^2$ and the fermions $\psi$ are attached to it, the vertices $x_{1,2,3}'$ of the triangle become
unique and can be integrated over, resulting in the conformal
three-point function
\begin{align}
\label{general three point function}
\langle s^2(x_1)\bar\psi (x_2)\psi (x_3)\rangle = 
C_{s^2\bar\psi\psi}\,\frac{\mu^{-2\gamma_\psi-\gamma_{s^2}}\,
x_{23}^\mu \gamma_\mu}{|x_{23}|^{d-2+2\gamma_\psi-\gamma_{s^2}}
(|x_{12}||x_{13}|)^{2+\gamma_{s^2}}}\,.
\end{align}
Here we have defined the amplitude coefficient
\begin{equation}
\label{Cs2psi psi general}
C_{s^2\bar\psi\psi}=-\frac{1}{N}\,C_{s^2}C_\psi^2\,(1+A_{s^2})(1+A_\psi)^2\,Z_{s^2\bar\psi\psi}\,{\cal U}\,,
\end{equation}
where $C_{s^2}$ is given by (\ref{Cs2 def}), and $A_{s^2,\psi}$ are amplitude corrections to the $s^2$,
$\psi$ propagators,
and the factor of ${\cal U}$ originates from application of the
uniqueness relations (\ref{scalar uniqueness}), (\ref{fermion uniqueness}) 
\begin{align}
{\cal U} &= -\pi^{d}\, U\left(2+\gamma_{s^2},\frac{d-\gamma_{s^2}}{2}-1,
\frac{d-\gamma_{s^2}}{2}-1\right)\,A\left(1+\frac{\gamma_{s^2}}{2}\right)A\left(\frac{d-\gamma_{s^2}}{2}-1\right)\notag\\
&\times V\left(\frac{d-1-\gamma_{s^2}}{2}-\gamma_{\psi},\frac{d-1}{2}+\gamma_\psi\right)
 V\left(\frac{d-1}{2}+\gamma_\psi,
\frac{3+\gamma_{s^2}}{2}-\gamma_\psi\right)\\
&=U_0\left(1+\delta u+{\cal O}\left(\frac{1}{N^2}\right)\right)\,.\notag
\end{align}
where
\begin{align}
\label{U0}
U_0&=-\frac{2 \pi ^{\frac{3 d}{2}}}{(d-4) \Gamma \left(\frac{d}{2}\right)^3}\,,\\
\delta u &=\frac{2  (d-4)^2\, \gamma _ \psi - (3 (d-8) d+40)\, \gamma _{s^2}}{2 (d-4) (d-2)}\,.
\label{delta u}
\end{align}

To separate out the leading and sub-leading contributions to the OPE
coefficient,
let us expand (\ref{Cs2psi psi general}) in $1/N$ as 
\begin{equation}
C_{s^2\bar\psi\psi} = C_{s^2\bar\psi\psi}^{(1/N)} (1+ \delta C_{s^2\bar\psi\psi})\,,
\end{equation}
At the leading order the three-point function $\langle s^2\bar\psi \psi \rangle$
is determined by the ${\cal O}(1/N)$ diagram
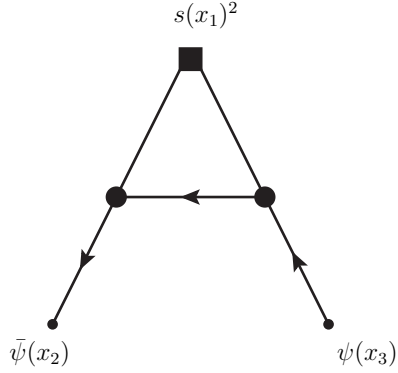
\begin{center}
  \begin{picture}(160,132) (59,15)
    \SetWidth{1.0}
    \SetColor{Black}
    \CBox(120,120)(128,128){Black}{Black}
    \Line[](120,120)(96,72)
    \Line[](128,120)(152,72)
    \Line[arrow,arrowpos=0.5,arrowlength=5,arrowwidth=2,arrowinset=0.2](152,72)(96,72)
    \Line[arrow,arrowpos=0.5,arrowlength=5,arrowwidth=2,arrowinset=0.2](96,72)(72,24)
    \Line[arrow,arrowpos=0.5,arrowlength=5,arrowwidth=2,arrowinset=0.2](176,24)(152,72)
    \Vertex(96,72){4}
    \Vertex(152,72){4}
    \Vertex(72,24){2}
    \Vertex(176,24){2}
    \Text(56,8)[lb]{\scalebox{0.8}{$\bar\psi(x_2)$}}
    \Text(180,8)[lb]{\scalebox{0.8}{$\psi(x_3)$}}
    \Text(118,136)[lb]{\scalebox{0.8}{$s(x_1)^2$}}
  \end{picture}
\end{center}
and was evaluated in \cite{Goykhman:2020ffn}
\begin{equation}
\label{leading order s^2 psi psi coefficient}
C_{s^2\bar\psi\psi}^{(1/N)}=\frac{1}{N}\,
\frac{2^{2 d-3} \pi ^{-\frac{d}{2}-3} (\cos (\pi  d)-1) \Gamma \left(\frac{d-1}{2}\right)^2}{\Gamma \left(\frac{d}{2}-1\right)}\,.
\end{equation}
Using this result we can determine the $s^2\bar\psi\psi$ conformal triangle amplitude
\begin{equation}
\label{expansion of Zs2psipsi def}
Z_{s^2\bar\psi\psi}=Z_{s^2\bar\psi\psi}^{(0)}(1+\delta Z_{s^2\bar\psi\psi})
\end{equation}
at the leading order in $1/N$ as
\begin{equation}
\label{Leading order Z}
Z_{s^2\bar\psi\psi}^{(0)}=-N\,\frac{C_{s^2\bar\psi\psi}^{(1/N)}}{C_{s^2}C_\psi^2\,U_0}=
\frac{1}{2\pi^d}\, \frac{4-d}{d-2}\,\Gamma \left(\frac{d}{2}\right)^2\,.
\end{equation}

We proceed to calculating the diagrams contributing to the conformal
triangle amplitude correction $\delta Z_{s^2\bar\psi\psi}$
at the next-to-leading ${\cal O}(1/N)$ order,
which we represent as a sum of four terms.
Diagrammatically, the conformal triangle up to the next-to-leading order in $1/N$ is represented by the
following equation:
\begin{center}
  \begin{picture}(362,70) (14,-11)
    \SetWidth{1.0}
    \SetColor{Black}
    \CBox(51,53)(55,57){Black}{Black}
    \Line[](53,57)(53,29)
    \GOval(53,42)(3,3)(0){0.882}
    \Vertex(53,29){3}
    \Line[](53,29)(39,4)
    \Vertex(39,4){3}
    \Line[arrow,arrowpos=0.5,arrowlength=3,arrowwidth=1,arrowinset=0.2](68,4)(39,4)
    \Line[](68,4)(53,29)
    \Vertex(68,4){3}
    \Line[arrow,arrowpos=0.2,arrowlength=3,arrowwidth=1,arrowinset=0.2](92,-9)(68,4)
    \Line[arrow,arrowpos=0.8,arrowlength=3,arrowwidth=1,arrowinset=0.2](92,-9)(68,4)
    \GOval(80,-3)(3,3)(0){0.882}
    \Line[arrow,arrowpos=0.25,arrowlength=3,arrowwidth=1,arrowinset=0.2](39,4)(15,-9)
    \Line[arrow,arrowpos=0.8,arrowlength=3,arrowwidth=1,arrowinset=0.2](39,4)(15,-9)
    \GOval(27,-2)(3,3)(0){0.882}
    \Vertex(15,-9){2}
    \Vertex(92,-9){2}
    \Text(31,17)[lb]{\scalebox{0.8}{$2a$}}
    \Text(64,17)[lb]{\scalebox{0.8}{$2a$}}
    \Text(48,-7)[lb]{\scalebox{0.8}{$2b$}}
    \Text(-30,0)[lb]{\scalebox{1}{$\frac{-Z_{s^2\bar\psi\psi}}{N\,\mu^{2\gamma_\psi+\gamma_{s^2}}}\cdot$}}
    \Text(102,20)[lb]{\scalebox{1}{$=$}}
    \CBox(140,53)(144,57){Black}{Black}
    \Line[arrow,arrowpos=0.7,arrowlength=3,arrowwidth=1,arrowinset=0.2](142,56)(118,-9)
    \Line[arrow,arrowpos=0.9,arrowlength=3,arrowwidth=1,arrowinset=0.2](142,56)(118,-9)
    \Line[arrow,arrowpos=0.1,arrowlength=3,arrowwidth=1,arrowinset=0.2](169,-9)(142,56)
    \Line[arrow,arrowpos=0.3,arrowlength=3,arrowwidth=1,arrowinset=0.2](169,-9)(142,56)
    \Vertex(118,-9){2}
    \Vertex(169,-9){2}
    \Line[arrow,arrowpos=0.3,arrowlength=3,arrowwidth=1,arrowinset=0.2](158,17)(128,17)
    \Line[arrow,arrowpos=0.8,arrowlength=3,arrowwidth=1,arrowinset=0.2](158,17)(128,17)
    \GOval(143,17)(3,3)(0){0.882}
    \GOval(128,17)(3,3)(0){0.882}
    \GOval(158,17)(3,3)(0){0.882}
    \GOval(123,4)(3,3)(0){0.882}
    \GOval(163,4)(3,3)(0){0.882}
    \GOval(134,33)(3,3)(0){0.882}
    \GOval(151,33)(3,3)(0){0.882}
    \Text(180,20)[lb]{\scalebox{1}{$+$}}
    \CBox(218,53)(222,57){Black}{Black}
    \Line[arrow,arrowpos=0.6,arrowlength=5,arrowwidth=2,arrowinset=0.2](220,56)(196,-9)
    \Line[arrow,arrowpos=0.92,arrowlength=5,arrowwidth=2,arrowinset=0.2](220,56)(196,-9)
    \Line[arrow,arrowpos=0.1,arrowlength=5,arrowwidth=2,arrowinset=0.2](248,-9)(220,56)
    \Line[arrow,arrowpos=0.43,arrowlength=5,arrowwidth=2,arrowinset=0.2](248,-9)(220,56)
    \Vertex(196,-9){2}
    \Vertex(248,-9){2}
    \Line[arrow,arrowpos=0.5,arrowlength=5,arrowwidth=2,arrowinset=0.2](232,30)(210,30)
    \Vertex(210,30){3}
    \Vertex(232,30){3}
    \Line[](202,4)(243,4)
    \Vertex(202,4){3}
    \Vertex(243,4){3}
    \Text(258,20)[lb]{\scalebox{1}{$+$}}
    \CBox(296,53)(300,57){Black}{Black}
    \Line[arrow,arrowpos=0.92,arrowlength=5,arrowwidth=2,arrowinset=0.2](298,56)(274,-9)
    \Line[arrow,arrowpos=0.42,arrowlength=5,arrowwidth=2,arrowinset=0.2](298,56)(274,-9)
    \Line[arrow,arrowpos=0.1,arrowlength=5,arrowwidth=2,arrowinset=0.2](326,-9)(298,56)
    \Line[arrow,arrowpos=0.60,arrowlength=5,arrowwidth=2,arrowinset=0.2](326,-9)(298,56)
    \Vertex(274,-9){2}
    \Vertex(326,-9){2}
    \Line[arrow,arrowpos=0.5,arrowlength=5,arrowwidth=2,arrowinset=0.2](321,5)(280,5)
    \Vertex(320,5){3}
    \Vertex(280,5){3}
    \Line[arrow,arrowpos=0.5,arrowlength=5,arrowwidth=2,arrowinset=0.2](286,21)(315,21)
    \Vertex(313,21){3}
    \Vertex(286,21){3}
    \Line[arrow,arrowpos=0.5,arrowlength=5,arrowwidth=2,arrowinset=0.2](307,36)(291,36)
    \Vertex(291,36){3}
    \Vertex(307,36){3}
    \Text(336,20)[lb]{\scalebox{1}{$+$}}
    \CBox(400,53)(404,57){Black}{Black}
    \Line[](387,20)(402,56)
    \Line[](369,36)(402,56)
    \Vertex(369,36){3}
    \Line[arrow,arrowpos=0.5,arrowlength=5,arrowwidth=2,arrowinset=0.2](386,36)(369,36)
    \Vertex(386,36){3}
    \Line[arrow,arrowpos=0.5,arrowlength=5,arrowwidth=2,arrowinset=0.2](386,20)(386,36)
    \Vertex(386,20){3}
    \Line[arrow,arrowpos=0.2,arrowlength=5,arrowwidth=2,arrowinset=0.2](369,36)(354,-9)
    \Line[arrow,arrowpos=0.84,arrowlength=5,arrowwidth=2,arrowinset=0.2](369,36)(354,-9)
    \Vertex(354,-9){2}
    \Line[arrow,arrowpos=0.17,arrowlength=5,arrowwidth=2,arrowinset=0.2](405,-9)(386,36)
    \Vertex(405,-9){2}
    \Line[arrow,arrowpos=0.5,arrowlength=5,arrowwidth=2,arrowinset=0.2](400,5)(360,5)
    \Vertex(399,5){3}
    \Vertex(359,5){3}
    \Line[arrow,arrowpos=0.5,arrowlength=5,arrowwidth=2,arrowinset=0.2](365,20)(387,20)
    \Vertex(365,20){3}
  \end{picture}
\end{center}
The gray blobs stand for the dressed $\psi$ and $s$ propagators and the dressed $s\bar\psi\psi$
vertex (represented by the corresponding  $s\bar\psi\psi$ conformal triangle), according to the conventions introduced in section~\ref{sec:setup}.
We have also assigned the exponents of $2a$ and $2b$ to the internal lines of the
$s^2\bar\psi\psi$  conformal
triangle, see (\ref{a and b for conformal triangle}).

A straightforward way to calculate the $\delta Z_{s^2\bar\psi\psi}$ is furnished by the background field
method.\footnote{See \cite{Goykhman:2020ffn} where the background field method was first applied
to obtain CFT data in vector models.} To this end we split the Hubbard-Stratonovich field into the non-dynamical 
background component $\bar s$ and the fluctuating component $s$, and isolate the  ${\cal O}(\bar s^2)$
terms quadratic in the background $\bar s$. The resulting diagrammatic equation describes
the fermionic propagator $\langle \psi\bar\psi  \rangle |_{\bar s}$ in the $\bar s$ background,
at the second order in $\bar s$:
\begin{center}
  \begin{picture}(362,55) (14,-11)
    \SetWidth{1.0}
    \SetColor{Black}
    \Text(50,35)[lb]{\scalebox{0.8}{$\bar s^2$}}
    \Vertex(53,29){3}
    \Line[](53,29)(39,4)
    \Vertex(39,4){3}
    \Line[arrow,arrowpos=0.5,arrowlength=3,arrowwidth=1,arrowinset=0.2](68,4)(39,4)
    \Line[](68,4)(53,29)
    \Vertex(68,4){3}
    \Line[arrow,arrowpos=0.2,arrowlength=3,arrowwidth=1,arrowinset=0.2](92,-9)(68,4)
    \Line[arrow,arrowpos=0.8,arrowlength=3,arrowwidth=1,arrowinset=0.2](92,-9)(68,4)
    \GOval(80,-3)(3,3)(0){0.882}
    \Line[arrow,arrowpos=0.25,arrowlength=3,arrowwidth=1,arrowinset=0.2](39,4)(15,-9)
    \Line[arrow,arrowpos=0.8,arrowlength=3,arrowwidth=1,arrowinset=0.2](39,4)(15,-9)
    \GOval(27,-2)(3,3)(0){0.882}
    \Vertex(15,-9){2}
    \Vertex(92,-9){2}
    \Text(31,17)[lb]{\scalebox{0.8}{$2a$}}
    \Text(64,17)[lb]{\scalebox{0.8}{$2a$}}
    \Text(48,-7)[lb]{\scalebox{0.8}{$2b$}}
    \Text(-30,0)[lb]{\scalebox{1}{$\frac{-Z_{s^2\bar\psi\psi}}{N\,\mu^{2\gamma_\psi-\gamma_{s^2}}}\cdot$}}
    \Text(102,0)[lb]{\scalebox{1}{$=$}}
    \Line[arrow,arrowpos=0.3,arrowlength=3,arrowwidth=1,arrowinset=0.2](128,17)(118,-9)
    \Line[arrow,arrowpos=0.8,arrowlength=3,arrowwidth=1,arrowinset=0.2](128,17)(118,-9)
    \Line[arrow,arrowpos=0.2,arrowlength=3,arrowwidth=1,arrowinset=0.2](169,-9)(158,17)
    \Line[arrow,arrowpos=0.8,arrowlength=3,arrowwidth=1,arrowinset=0.2](169,-9)(158,17)
    \Vertex(118,-9){2}
    \Vertex(169,-9){2}
    \Line[arrow,arrowpos=0.3,arrowlength=3,arrowwidth=1,arrowinset=0.2](158,17)(128,17)
    \Line[arrow,arrowpos=0.8,arrowlength=3,arrowwidth=1,arrowinset=0.2](158,17)(128,17)
    \GOval(143,17)(3,3)(0){0.882}
    \GOval(128,17)(3,3)(0){0.882}
    \GOval(158,17)(3,3)(0){0.882}
    \GOval(123,4)(3,3)(0){0.882}
    \GOval(163,4)(3,3)(0){0.882}
    \Text(180,0)[lb]{\scalebox{1}{$+$}}
    \Line[arrow,arrowpos=0.35,arrowlength=5,arrowwidth=2,arrowinset=0.2](210,30)(196,-9)
    \Line[arrow,arrowpos=0.85,arrowlength=5,arrowwidth=2,arrowinset=0.2](210,30)(196,-9)
    \Line[arrow,arrowpos=0.2,arrowlength=5,arrowwidth=2,arrowinset=0.2](248,-9)(232,30)
    \Line[arrow,arrowpos=0.7,arrowlength=5,arrowwidth=2,arrowinset=0.2](248,-9)(232,30)
    \Vertex(196,-9){2}
    \Vertex(248,-9){2}
    \Line[arrow,arrowpos=0.5,arrowlength=5,arrowwidth=2,arrowinset=0.2](232,30)(210,30)
    \Vertex(210,30){3}
    \Vertex(232,30){3}
    \Line[](202,4)(243,4)
    \Vertex(202,4){3}
    \Vertex(243,4){3}
    \Text(258,0)[lb]{\scalebox{1}{$+$}}
    \Line[arrow,arrowpos=0.85,arrowlength=5,arrowwidth=2,arrowinset=0.2](291,36)(274,-9)
    \Line[arrow,arrowpos=0.2,arrowlength=5,arrowwidth=2,arrowinset=0.2](291,36)(274,-9)
    \Line[arrow,arrowpos=0.15,arrowlength=5,arrowwidth=2,arrowinset=0.2](326,-9)(307,36)
    \Line[arrow,arrowpos=0.83,arrowlength=5,arrowwidth=2,arrowinset=0.2](326,-9)(307,36)
    \Vertex(274,-9){2}
    \Vertex(326,-9){2}
    \Line[arrow,arrowpos=0.5,arrowlength=5,arrowwidth=2,arrowinset=0.2](321,5)(280,5)
    \Vertex(320,5){3}
    \Vertex(280,5){3}
    \Line[arrow,arrowpos=0.5,arrowlength=5,arrowwidth=2,arrowinset=0.2](286,21)(315,21)
    \Vertex(313,21){3}
    \Vertex(286,21){3}
    \Line[arrow,arrowpos=0.5,arrowlength=5,arrowwidth=2,arrowinset=0.2](307,36)(291,36)
    \Vertex(291,36){3}
    \Vertex(307,36){3}
    \Text(336,0)[lb]{\scalebox{1}{$+$}}
    \Vertex(369,36){3}
    \Line[arrow,arrowpos=0.5,arrowlength=5,arrowwidth=2,arrowinset=0.2](386,36)(369,36)
    \Vertex(386,36){3}
    \Line[arrow,arrowpos=0.5,arrowlength=5,arrowwidth=2,arrowinset=0.2](386,20)(386,36)
    \Vertex(386,20){3}
    \Line[arrow,arrowpos=0.2,arrowlength=5,arrowwidth=2,arrowinset=0.2](369,36)(354,-9)
    \Line[arrow,arrowpos=0.84,arrowlength=5,arrowwidth=2,arrowinset=0.2](369,36)(354,-9)
    \Vertex(354,-9){2}
    \Line[arrow,arrowpos=0.17,arrowlength=5,arrowwidth=2,arrowinset=0.2](405,-9)(386,36)
    \Vertex(405,-9){2}
    \Line[arrow,arrowpos=0.5,arrowlength=5,arrowwidth=2,arrowinset=0.2](400,5)(360,5)
    \Vertex(399,5){3}
    \Vertex(359,5){3}
    \Line[arrow,arrowpos=0.5,arrowlength=5,arrowwidth=2,arrowinset=0.2](365,20)(387,20)
    \Vertex(365,20){3}
  \end{picture}
\end{center}
On the l.h.s. of this equation we obtain
\begin{center}
  \begin{picture}(362,65) (14,-26)
    \SetWidth{1.0}
    \SetColor{Black}
    \Text(50,35)[lb]{\scalebox{0.8}{$\bar s^2$}}
    \Vertex(53,29){3}
    \Line[](53,29)(39,4)
    \Vertex(39,4){3}
    \Line[arrow,arrowpos=0.5,arrowlength=3,arrowwidth=1,arrowinset=0.2](68,4)(39,4)
    \Line[](68,4)(53,29)
    \Vertex(68,4){3}
    \Line[arrow,arrowpos=0.2,arrowlength=3,arrowwidth=1,arrowinset=0.2](92,-9)(68,4)
    \Line[arrow,arrowpos=0.8,arrowlength=3,arrowwidth=1,arrowinset=0.2](92,-9)(68,4)
    \GOval(80,-3)(3,3)(0){0.882}
    \Line[arrow,arrowpos=0.25,arrowlength=3,arrowwidth=1,arrowinset=0.2](39,4)(15,-9)
    \Line[arrow,arrowpos=0.8,arrowlength=3,arrowwidth=1,arrowinset=0.2](39,4)(15,-9)
    \GOval(27,-2)(3,3)(0){0.882}
    \Vertex(15,-9){2}
    \Vertex(92,-9){2}
    \Text(31,17)[lb]{\scalebox{0.8}{$2a$}}
    \Text(64,17)[lb]{\scalebox{0.8}{$2a$}}
    \Text(48,-7)[lb]{\scalebox{0.8}{$2b$}}
    \Text(5,-25)[lb]{\scalebox{0.8}{$\bar\psi(x_2)$}}
    \Text(88,-25)[lb]{\scalebox{0.8}{$\psi(x_3)$}}
    \Text(-30,0)[lb]{\scalebox{1}{$\frac{-Z_{s^2\bar\psi\psi}}{N\,\mu^{2\gamma_\psi-\gamma_{s^2}}}\cdot$}}
    \Text(102,-5)[lb]{\scalebox{1}{$=\langle\psi (x_3)\bar\psi(x_2)\rangle \Bigg|_{\bar s}=
    \frac{C_{s^2\bar\psi\psi}^{(1/N)}}{C_{s^2}}
    \,(1{+}\delta u {+}\delta Z_{s^2\bar\psi\psi}{+}2A_\psi)
    \frac{\mu^{\gamma_{s^2}-2\gamma_\psi}\,x_{23}^\mu\gamma_\mu}{|x_{23}|^{d-2+2\gamma_\psi-\gamma_{s^2}}}$}}
  \end{picture}
\end{center}
where we took into account (\ref{U0}), (\ref{delta u}), (\ref{Leading order Z}), used the 
uniqueness relation (\ref{fermion uniqueness}), and linearized over the $1/N$ corrections
contributing to the overall amplitude.

We proceed to calculating the diagrams on the r.h.s. of the diagrammatic equation
for the $\langle \psi\bar\psi\rangle _{\bar s}$.
The first contribution is due to 
the leading order $\langle s^2\bar\psi \psi \rangle$ diagram in which all the
internal $s\bar\psi\psi$ vertices and $\psi$, $s$ propagators have been dressed,
and $s^2$ was set to the background value $\bar s^2$:
\begin{center}
  \begin{picture}(400,70) (59,15)
    \SetWidth{1.0}
    \SetColor{Black}
    \Line[arrow,arrowpos=0.25,arrowlength=3,arrowwidth=2,arrowinset=0.2](152,72)(96,72)
    \Line[arrow,arrowpos=0.75,arrowlength=3,arrowwidth=2,arrowinset=0.2](152,72)(96,72)
    \Line[arrow,arrowpos=0.25,arrowlength=3,arrowwidth=2,arrowinset=0.2](96,72)(72,24)
    \Line[arrow,arrowpos=0.8,arrowlength=3,arrowwidth=2,arrowinset=0.2](96,72)(72,24)
    \Line[arrow,arrowpos=0.2,arrowlength=3,arrowwidth=2,arrowinset=0.2](176,24)(152,72)
    \Line[arrow,arrowpos=0.75,arrowlength=3,arrowwidth=2,arrowinset=0.2](176,24)(152,72)
    \Vertex(72,24){2}
    \Vertex(176,24){2}
    \GOval(96,72)(8,8)(0){0.882}
    \GOval(125,72)(8,8)(0){0.882}
    \GOval(151,72)(8,8)(0){0.882}
    \GOval(84,47)(8,8)(0){0.882}
    \GOval(164,47)(8,8)(0){0.882}
    \Text(56,8)[lb]{\scalebox{0.8}{$\bar\psi(x_2)$}}
    \Text(176,8)[lb]{\scalebox{0.8}{$\psi(x_3)$}}
    \Text(200,40)[lb]{\scalebox{1}{$=\frac{C_{s^2\bar\psi\psi}^{(1/N)}}{C_{s^2}}
    (1+ 3A_\psi +2\delta Z_{s\bar\psi\psi} + W_1)
    \,\frac{\mu^{2\gamma_{s}-2\gamma_\psi}\,
    x_{23}^\mu \gamma_\mu}{|x_{23}|^{d-2+2\gamma_\psi-2\gamma_{s}}}$}}
  \end{picture}
\end{center}
Notice that this diagram contains the entire leading order contribution to the $\langle \psi\bar\psi\rangle |_{\bar s}$
two-point function.
We linearized over the next-to-leading in $1/N$ terms, obtaining the sum of $3A_\psi$ due to
amplitude corrections to three internal dressed fermion propagators and $2\delta Z_{s\bar\psi \psi}$
due to two dressed $s\bar\psi\psi$ vertices. Finally, $W_1$ is obtained from the equation
\begin{align}
&1+W_1=\frac{C_{s^2}}{C_{s^2\bar\psi\psi}^{(1/N)}}
\left(-\frac{Z_{s\bar\psi\psi}^{(0)}}{\sqrt{N}}\right)^2C_\psi^3
(-\pi^{2d})V\left(\frac{d-1}{2}+\gamma_\psi,\frac{d-1-\gamma_s}{2}\right)^3
A\left(1{-}\gamma_\psi{+}\frac{\gamma_s}{2}\right)^3\notag\\
&\times U\left(1{+}\gamma_s,\frac{d{-}\gamma_s}{2}{-}\gamma_\psi,\frac{d{-}\gamma_s}{2}{-}1{+}\gamma_\psi\right)^2A\left(\frac{d{-}\gamma_s}{2}{-}1{+}\gamma_\psi\right)
V\left(\frac{d{-}1{-}\gamma_s}{2},\frac{3}{2}{-}\gamma_\psi{+}\gamma_s\right)\notag\,,
\end{align}
resulting from a repeated application of the uniqueness and the propagator merging relations, and
expanded to the next-to-leading order in $1/N$.
Further expanding the r.h.s. in $\gamma_{\psi, s}={\cal O}(1/N)$ we obtain
\begin{align}
\label{W1}
W_1 = \frac{d-4}{d-2}\,\gamma_\psi+\frac{8-d}{d-2}\,\gamma_s\,.
\end{align}

Next, we consider the contribution represented by the diagram
\begin{center}
  \begin{picture}(400,87) (59,5)
    \SetWidth{1.0}
    \SetColor{Black}
    \Line[arrow,arrowpos=0.3,arrowlength=5,arrowwidth=2,arrowinset=0.2](104,88)(72,24)
    \Line[arrow,arrowpos=0.82,arrowlength=5,arrowwidth=2,arrowinset=0.2](104,88)(72,24)
    \Line[arrow,arrowpos=0.7,arrowlength=5,arrowwidth=2,arrowinset=0.2](176,24)(144,88)
    \Line[arrow,arrowpos=0.2,arrowlength=5,arrowwidth=2,arrowinset=0.2](176,24)(144,88)
    \Text(56,8)[lb]{\scalebox{0.8}{$\bar\psi(x_2)$}}
    \Text(176,8)[lb]{\scalebox{0.8}{$\psi(x_3)$}}
    \Line[arrow,arrowpos=0.5,arrowlength=5,arrowwidth=2,arrowinset=0.2](144,88)(104,88)
    \Line[](85,48)(165,48)
    \Vertex(85,48){4}
    \Vertex(165,48){4}
    \Vertex(104,88){4}
    \Vertex(144,88){4}
    \Vertex(72,24){2}
    \Vertex(176,24){2}
    \Text(220,40)[lb]{\scalebox{1}{$=\frac{C_{s^2\bar\psi\psi}^{(1/N)}}{C_{s^2}} \, W_2
    \,\frac{x_{23}^\mu \gamma_\mu}{|x_{23}|^{d-2}}$}}
  \end{picture}
\end{center}
Repeated application of the propagator merging relations gives
\begin{align}
W_2&=\frac{C_{s^2}}{C_{s^2\bar\psi\psi}^{(1/N)}}\left({-}\frac{1}{\sqrt{N}}\right)^4
C_\psi^5 C_s\pi^{2d}A(1)^2V\left(\frac{d{-}1}{2},\frac{d{-}1}{2}\right)^2
A\left(\frac{d}{2}{-}1\right)^2V\left(\frac{d{-}1}{2},\frac{3}{2}\right)^2\notag\\
&=\frac{1}{N}\,\frac{2^{d-2} \sin \left(\frac{\pi  d}{2}\right) \Gamma \left(\frac{d-1}{2}\right)}{\pi ^{3/2} \Gamma \left(\frac{d}{2}\right)}\,.
\end{align}

The remaining contributions $W_{3,4}$ to the conformal triangle amplitude $\delta Z_{s^2\bar\psi\psi}$
originate from the next-to-leading corrections to the $\langle s^2 s s\rangle$ sub-diagram
of the leading order $\langle s^2\bar\psi\psi\rangle$ diagram. 
To regularize these divergent diagrams we add a small correction $\delta/2$
to the propagators of the internal $s$ lines, following the technique reviewed in
section~\ref{sec:setup}:\footnote{The choice of $\delta/2$ instead of $\delta$ is a convention to make the total
power of the diagram add up to $\delta$.}
\begin{center}
  \begin{picture}(400,97) (59,5)
    \SetWidth{1.0}
    \SetColor{Black}
    \Line[arrow,arrowpos=0.83,arrowlength=5,arrowwidth=2,arrowinset=0.2](72,24)(108,96)
    \Line[arrow,arrowpos=0.88,arrowlength=5,arrowwidth=2,arrowinset=0.2](108,96)(72,24)
    \Line[arrow,arrowpos=0.20,arrowlength=5,arrowwidth=2,arrowinset=0.2](140,96)(176,24)
    \Line[arrow,arrowpos=0.15,arrowlength=5,arrowwidth=2,arrowinset=0.2](176,24)(140,96)
    \Text(56,8)[lb]{\scalebox{0.8}{$\bar\psi(x_2)$}}
    \Text(176,8)[lb]{\scalebox{0.8}{$\psi(x_3)$}}
    \Vertex(72,24){2}
    \Vertex(176,24){2}
    \Line[arrow,arrowpos=0.5,arrowlength=5,arrowwidth=2,arrowinset=0.2](152,72)(96,72)
    \Line[arrow,arrowpos=0.5,arrowlength=5,arrowwidth=2,arrowinset=0.2](163,48)(85,48)
    \Line[arrow,arrowpos=0.5,arrowlength=5,arrowwidth=2,arrowinset=0.2](108,96)(140,96)
    \Vertex(108,96){4}
    \Vertex(140,96){4}
    \Vertex(152,72){4}
    \Vertex(96,72){4}
    \Vertex(85,48){4}
    \Vertex(163,48){4}
    \Text(55,58)[lb]{\scalebox{0.8}{$2+\delta/2$}}
    \Text(165,58)[lb]{\scalebox{0.8}{$2+\delta/2$}}
    \Text(216,50)[lb]{\scalebox{1}{$=\frac{C_{s^2\bar\psi\psi}^{(1/N)}}{C_{s^2}} \, W_3
    \,\frac{x_{23}^\mu \gamma_\mu\,\mu^{-\delta}}{|x_{23}|^{d-2+\delta}}$}}
  \end{picture}
\end{center}
\begin{center}
  \begin{picture}(400,97) (59,5)
    \SetWidth{1.0}
    \SetColor{Black}
    \Line[arrow,arrowpos=0.83,arrowlength=5,arrowwidth=2,arrowinset=0.2](72,24)(108,96)
    \Line[arrow,arrowpos=0.88,arrowlength=5,arrowwidth=2,arrowinset=0.2](108,96)(72,24)
    \Text(56,8)[lb]{\scalebox{0.8}{$\bar\psi(x_2)$}}
    \Text(176,8)[lb]{\scalebox{0.8}{$\psi(x_3)$}}
    \Vertex(72,24){2}
    \Vertex(176,24){2}
    \Line[arrow,arrowpos=0.5,arrowlength=5,arrowwidth=2,arrowinset=0.2](165,48)(88,48)
    \Line[arrow,arrowpos=0.5,arrowlength=5,arrowwidth=2,arrowinset=0.2](112,96)(144,96)
    \Vertex(108,96){4}
    \Vertex(144,96){4}
    \Vertex(144,72){4}
    \Vertex(96,72){4}
    \Vertex(85,48){4}
    \Vertex(165,48){4}
    \Text(55,58)[lb]{\scalebox{0.8}{$2+\delta/2$}}
    \Text(162,63)[lb]{\scalebox{0.8}{$2+\delta/2$}}
    \Line[arrow,arrowpos=0.15,arrowlength=5,arrowwidth=2,arrowinset=0.2](176,24)(144,96)
    \Line[arrow,arrowpos=0.5,arrowlength=5,arrowwidth=2,arrowinset=0.2](144,96)(144,72)
    \Line[arrow,arrowpos=0.5,arrowlength=5,arrowwidth=2,arrowinset=0.2](144,72)(96,72)
    \Text(216,50)[lb]{\scalebox{1}{$=\frac{C_{s^2\bar\psi\psi}^{(1/N)}}{C_{s^2}} \, W_4
    \,\frac{x_{23}^\mu \gamma_\mu\,\mu^{-\delta}}{|x_{23}|^{d-2+\delta}}$}}
  \end{picture}
\end{center}
Here we have\footnote{The overall minus sign is due to the Feynman rule for the fermion loop, see section~\ref{sec:setup}, while $2$ is the factor of symmetry.}
\begin{align}
W_3&=-\frac{C_{s^2}}{C_{s^2\bar\psi\psi}^{(1/N)}}\,
2N\,\left(-\frac{1}{\sqrt{N}}\right)^6  C_\psi^7C_s^2 (-\pi^{2d})A(1)V\left(\frac{d-1}{2},\frac{d-1}{2}\right)
A\left(\frac{d}{2}-1\right)\notag\\
&\times V\left(\frac{d-1}{2},\frac{3}{2}\right)U\left(1+\frac{\delta}{4},1-\frac{\delta}{4},d-2\right)
U\left(1+\frac{\delta}{4},\frac{2d+\delta}{4}-1,\frac{d-\delta}{2}\right)\notag\\
&\times A\left(1-\frac{\delta}{2}\right)V\left(\frac{d-1+\delta}{2},\frac{d-1}{2}\right)A\left(\frac{d+\delta}{2}-1\right)
V\left(\frac{d-1}{2},\frac{3-\delta}{2}\right)\notag\\
&=\frac{1}{N}\,\frac{2 (d-2) \sin \left(\frac{\pi  d}{2}\right) \Gamma (d-1)}{\pi  \Gamma \left(\frac{d}{2}\right)^2}\,
\left(\frac{2}{\delta} + 1\right)\,.\label{W3}
\end{align}
and
\begin{align}
\label{W4}
W_4 = \frac{1}{d-2}\,W_3\,.
\end{align}

Finally, we notice that the divergent terms of the last two diagrams add up to
\begin{equation}
W_3+ W_4 \supset \frac{2\gamma_s-\gamma_{s^2}}{\delta}\,,
\end{equation}
which is cancelled by the counter-term diagram
\begin{center}
  \begin{picture}(120,72) (59,15)
    \SetWidth{1.0}
    \SetColor{Black}
    \Line[arrow,arrowpos=0.5,arrowlength=5,arrowwidth=2,arrowinset=0.2](152,72)(96,72)
    \Line[arrow,arrowpos=0.5,arrowlength=5,arrowwidth=2,arrowinset=0.2](96,72)(72,24)
    \Line[arrow,arrowpos=0.5,arrowlength=5,arrowwidth=2,arrowinset=0.2](176,24)(152,72)
    \Vertex(96,72){4}
    \Vertex(152,72){4}
    \Vertex(72,24){2}
    \Vertex(176,24){2}
    \Text(-20,48)[lb]{\scalebox{1}{$\frac{\gamma_{s^2}-2\gamma_s}{\delta}\times$}}
    \Text(56,8)[lb]{\scalebox{0.8}{$\bar\psi(x_2)$}}
    \Text(180,8)[lb]{\scalebox{0.8}{$\psi(x_3)$}}
  \end{picture}
\end{center}
induced by the renormalization of the Hubbard-Stratonovich field $s$
and the composite operator $s^2$
\begin{equation}
\label{s and s2 renormalization}
s\rightarrow \sqrt{1+\frac{2\gamma_s}{\delta}}\,s\,,\qquad
s^2\rightarrow \sqrt{1+\frac{2\gamma_{s^2}}{\delta}}\,s^2\,.
\end{equation}

To summarize, we arrive at the following expression
\begin{equation}
\label{delta Z s2psi psi result}
\boxed{
\delta Z_{s^2\bar\psi\psi} = \sum_{i=1}^4 W_i\Bigg|_{\textrm{finite}} +2\delta Z_{s\bar\psi\psi} + A_\psi -\delta u
}
\end{equation}
In what follows we will skip specifying explicitly that only the finite part of the sum of $W_i$
is retained, implying that the infinities have been cancelled out by the counter-term.

\subsection{$s^2 s s$ conformal triangle}
\label{subsec:s2ss}

In section~\ref{subsec:s2psipsi} we demonstrated how the
 $\langle s^2\bar\psi\psi\rangle$ three-point
function can be expressed in terms of the corresponding $s^2\bar\psi\psi$ conformal
triangle (\ref{Cs2psi psi general}),
and proceeded to calculate the latter, arriving at (\ref{delta Z s2psi psi result}).
The results of the previous section can in fact be used
to calculate the $ s^2 s s$ conformal triangle at the next-to-leading
order in $1/N$, which we will use below to derive the $\langle s^2 s s\rangle$ and $\langle s^2 s^2\rangle$
correlation functions.

To this end, consider the dressed $s^2 s s$ vertex expressed in terms of the corresponding conformal triangle
\begin{center}
  \begin{picture}(200,68) (-20,-5)
    \SetWidth{1.0}
    \SetColor{Black}
    \GOval(-17,24)(14,14)(0){0.882}
    \Line[](-17,65)(-17,38)
    \Line[](-26,12)(-42,-7)
    \Line[](8,-7)(-8,12)
    \Line[](173,38)(145,4)
    \Vertex(145,4){4}
    \Line[](198,4)(173,38)
    \Line[](198,4)(145,4)
    \Vertex(198,4){4}
    \Vertex(173,38){4}
    \Text(30,20)[lb]{\scalebox{1}{$=-Z_{s^2ss}\,\mu^{\gamma_{s^2}+2\gamma_s}\times$}}
    \Text(143,20)[lb]{\scalebox{0.8}{$2a'$}}
    \Text(190,20)[lb]{\scalebox{0.8}{$2a'$}}
    \Text(165,-10)[lb]{\scalebox{0.8}{$2b'$}}
    \Text(178,43)[lb]{\scalebox{0.8}{$s(x_1)^2$}}
    \Text(133,-16)[lb]{\scalebox{0.8}{$s(x_2)$}}
    \Text(190,-16)[lb]{\scalebox{0.8}{$s(x_3)$}}
  \end{picture}
\end{center}
where we denoted
\begin{equation}
a' = \frac{d-\gamma_{s^2}}{2} - 1\,,\qquad b' = \frac{d+\gamma_{s^2}}{2} - \gamma_s\,,
\end{equation}
chosen so that when the $s^2$ and $s$ legs are attached to the triangle, its three vertices
become unique and can be integrated over, resulting in
\begin{center}
  \begin{picture}(100,98) (19,-20)
    \SetWidth{1.0}
    \SetColor{Black}
    \GOval(23,24)(14,14)(0){0.882}
    \Line[](23,65)(23,38)
    \Line[](14,12)(-2,-7)
    \Line[](48,-7)(32,12)
    \Vertex(-2,-7){2}
    \Vertex(48,-7){2}
    \CBox(20,62)(27,69){Black}{Black}
    \Text(20,72)[lb]{\scalebox{0.8}{$s(x_1)^2$}}
    \Text(-12,-22)[lb]{\scalebox{0.8}{$s(x_2)$}}
    \Text(43,-22)[lb]{\scalebox{0.8}{$s(x_3)$}}
    \GOval(40,3)(6,6)(0){0.882}
    \GOval(7,3)(6,6)(0){0.882}
    \GOval(23,50)(6,6)(0){0.882}
    \Text(100,20)[lb]{\scalebox{1}{$=\langle s(x_1)^2s(x_2)s(x_3)\rangle$}}
  \end{picture}
\end{center}
Conformal symmetry constraints the form of the $\langle s^2 s s\rangle$ three-point to be
\begin{equation}
\label{general <s^2 s s>}
\langle s(x_1)^2 s(x_2)s(x_3)\rangle = \mu^{-\gamma_{s^2}-2\gamma_s}\,\frac{C_{s^2ss}^{(0)}\left(1+\delta  C_{s^2ss}\right)}
{(|x_{12}||x_{13}|)^{2+\gamma_{s^2}}|x_{23}|^{2\gamma_s-\gamma_{s^2}}}\,,
\end{equation}
where
\begin{equation}
\label{leading order amplitude}
C_{s^2 ss}^{(0)} = 2C_s^2
\end{equation}
is the leading order OPE coefficient, $\delta  C_{s^2ss}$ stands for the sub-leading corrections.
The latter can be decomposed as
\begin{equation}
\label{delta Cs2ss in terms of delta V}
\delta  C_{s^2ss} = \delta V_{s^2 s s} + A_{s^2} + 2A_s\,,
\end{equation}
where the vertex amplitude correction $\delta V_{s^2 s s}$ is defined by the diagram
\begin{center}
  \begin{picture}(100,105) (19,-20)
    \SetWidth{1.0}
    \SetColor{Black}
    \GOval(23,24)(14,14)(0){0.882}
    \Line[](23,65)(23,38)
    \Line[](14,12)(-2,-7)
    \Line[](48,-7)(32,12)
    \Vertex(-2,-7){2}
    \Vertex(48,-7){2}
    \CBox(20,62)(27,69){Black}{Black}
    \Text(20,72)[lb]{\scalebox{0.8}{$s(x_1)^2$}}
    \Text(-12,-22)[lb]{\scalebox{0.8}{$s(x_2)$}}
    \Text(43,-22)[lb]{\scalebox{0.8}{$s(x_3)$}}
    \Text(46,0)[lb]{\scalebox{0.8}{$2+2\gamma_s$}}
    \Text(-30,0)[lb]{\scalebox{0.8}{$2+2\gamma_s$}}
    \Text(27,47)[lb]{\scalebox{0.8}{$4+2\gamma_{s^2}$}}
    \Text(100,20)[lb]{\scalebox{1}{$=\frac{C_{s^2 s s}^{(0)}\,(1+\delta V_{s^2ss})}{(|x_{12}||x_{13}|)^{2+\gamma_{s^2}}
    |x_{23}|^{2\gamma_s-\gamma_{s^2}}}$}}
  \end{picture}
\end{center}

Using the uniqueness to integrate over three vertices of the conformal triangle we can obtain
the relation between $C_{s^2 s s}$ and $Z_{s^2 s s}$:
\begin{align}
&C_{s^2ss}^{(0)}\,(1+\delta  V_{s^2ss})=-Z_{s^2ss}\,C_s^2C_{s^2}
U\left(\frac{d-\gamma_{s^2}}{2} - 1,\frac{d-\gamma_{s^2}}{2} - 1,2+\gamma_{s^2}\right) \\
&\times U\left(1+\frac{\gamma_{s^2}}{2},1+\gamma_s,d-2-\gamma_s-\frac{\gamma_{s^2}}{2}\right)
U\left(1+\gamma_s,\frac{d+\gamma_{s^2}}{2}-\gamma_s,\frac{d-\gamma_{s^2}}{2}-1\right)\,.\notag
\end{align}
Expanding
\begin{equation}
Z_{s^2ss} = Z_{s^2ss}^{(0)}(1+\delta Z_{s^2ss})
\end{equation}
we obtain
\begin{align}
Z_{s^2ss}^{(0)} &=\frac{ (d-4) \pi ^{3-\frac{3 d}{2}} \Gamma \left(\frac{d}{2}-1\right)^2 \Gamma (d-1)}
{2^{2 d+3}\,\Gamma \left(2-\frac{d}{2}\right) \Gamma \left(\frac{d-1}{2}\right)^2\, \sin ^2\left(\frac{\pi  d}{2}\right)}\,
(\gamma _{s^2}-2 \gamma _s) \,,\\
\delta Z_{s^2ss} &= \delta  V_{s^2ss} -\delta z\,,\label{delta Zs2ss in terms of delta Vs2ss}
\end{align}
where\footnote{Here $\gamma$ is the Euler constant, and we will denote the $n$th
derivative of the digamma function $\psi^{(0)}(x)=\Gamma'(x)/\Gamma(x)$ as $\psi^{(n)}(x)$.}
\begin{align}
\label{delta z expression}
\delta z&=\left(\frac{2}{d-2}+\pi  \cot \left(\frac{\pi  d}{2}\right)+\psi ^{(0)}(d-2)+\gamma\right)\,\gamma_s\\
&+\frac{1}{2} \left(\frac{4}{d-4}-\frac{2}{d-2}+\pi  \cot \left(\frac{\pi  d}{2}\right)+\psi ^{(0)}(d-2)+\gamma -2\right)\,\gamma_{s^2}\,.
\end{align}

We can use the last two diagrams contributing to the $\langle \bar\psi\psi\rangle|_{\bar s}$
correlation function, obtained in section~\ref{subsec:s2psipsi},
to extract the next-to-leading order correction $\delta Z_{s^2ss}$
to the $s^2ss$ conformal triangle.
The total of these diagrams is given by
\begin{center}
  \begin{picture}(400,100) (80,10)
    \SetWidth{1.0}
    \SetColor{Black}
    \Line[arrow,arrowpos=0.82,arrowlength=5,arrowwidth=2,arrowinset=0.2](72,24)(108,96)
    \Line[arrow,arrowpos=0.85,arrowlength=5,arrowwidth=2,arrowinset=0.2](108,96)(72,24)
    \Line[arrow,arrowpos=0.2,arrowlength=5,arrowwidth=2,arrowinset=0.2](140,96)(176,24)
    \Line[arrow,arrowpos=0.17,arrowlength=5,arrowwidth=2,arrowinset=0.2](176,24)(140,96)
    \Text(56,8)[lb]{\scalebox{0.8}{$\bar\psi(x_2)$}}
    \Text(176,8)[lb]{\scalebox{0.8}{$\psi(x_3)$}}
    \Vertex(72,24){2}
    \Vertex(176,24){2}
    \Line[arrow,arrowpos=0.5,arrowlength=5,arrowwidth=2,arrowinset=0.2](152,72)(96,72)
    \Line[arrow,arrowpos=0.5,arrowlength=5,arrowwidth=2,arrowinset=0.2](163,48)(85,48)
    \Line[arrow,arrowpos=0.5,arrowlength=5,arrowwidth=2,arrowinset=0.2](108,96)(140,96)
    \Vertex(108,96){4}
    \Vertex(140,96){4}
    \Vertex(152,72){4}
    \Vertex(96,72){4}
    \Vertex(85,48){4}
    \Vertex(163,48){4}
    \Text(55,58)[lb]{\scalebox{0.8}{$2+\delta/2$}}
    \Text(165,58)[lb]{\scalebox{0.8}{$2+\delta/2$}}
    \Line[arrow,arrowpos=0.83,arrowlength=5,arrowwidth=2,arrowinset=0.2](222,24)(258,96)
    \Line[arrow,arrowpos=0.85,arrowlength=5,arrowwidth=2,arrowinset=0.2](258,96)(222,24)
    \Text(206,8)[lb]{\scalebox{0.8}{$\bar\psi(x_2)$}}
    \Text(326,8)[lb]{\scalebox{0.8}{$\psi(x_3)$}}
    \Vertex(222,24){2}
    \Vertex(326,24){2}
    \Line[arrow,arrowpos=0.5,arrowlength=5,arrowwidth=2,arrowinset=0.2](315,48)(238,48)
    \Line[arrow,arrowpos=0.5,arrowlength=5,arrowwidth=2,arrowinset=0.2](262,96)(294,96)
    \Vertex(258,96){4}
    \Vertex(294,96){4}
    \Vertex(294,72){4}
    \Vertex(246,72){4}
    \Vertex(235,48){4}
    \Vertex(315,48){4}
    \Text(205,58)[lb]{\scalebox{0.8}{$2+\delta/2$}}
    \Text(312,63)[lb]{\scalebox{0.8}{$2+\delta/2$}}
    \Line[arrow,arrowpos=0.15,arrowlength=5,arrowwidth=2,arrowinset=0.2](326,24)(294,96)
    \Line[arrow,arrowpos=0.5,arrowlength=5,arrowwidth=2,arrowinset=0.2](294,96)(294,72)
    \Line[arrow,arrowpos=0.5,arrowlength=5,arrowwidth=2,arrowinset=0.2](294,72)(246,72)
    \Text(197,38)[lb]{\scalebox{0.8}{$+$}}
    \Text(340,30)[lb]{\scalebox{1}{$= \frac{C_{s^2\bar\psi\psi}^{(1/N)}}{C_{s^2}}
    \frac{(1+W_3+W_4)
    x_{23}^\mu \gamma_\mu}{\mu^{2\gamma_s-\gamma_{s^2}}\,|x_{23}|^{d-2+2\gamma_s-\gamma_{s^2}}}\Bigg|_{\frac{1}{N^2}}$}}
  \end{picture}
\end{center}
where in the r.h.s. of this equation we only retain the sub-leading terms due to the amplitude
$W_3+W_4$ and the anomalous dimensions $\gamma_{s,s^2}$. \footnote{Linearizing over the $1/N$ corrections,
one can explicitly verify that the total of the four diagrams, considered
in section~\ref{subsec:s2psipsi}, contributing to the correlation
function $\langle \bar\psi\psi\rangle |_{\bar s}$ at the next-to-leading order has the
conformal form.}
On the other hand, we can rewrite these diagrams using the dressed $s^2 s s$ vertex defined above:
\begin{center}
  \begin{picture}(90,80) (59,10)
    \SetWidth{1.0}
    \SetColor{Black}
    \Vertex(-98,24){2}
    \Vertex(6,24){2}
    \Text(-114,8)[lb]{\scalebox{0.8}{$\bar\psi(x_2)$}}
    \Text(10,8)[lb]{\scalebox{0.8}{$\psi(x_3)$}}
    \GOval(-46,80)(10,10)(0){0.882}
    \Line[arrow,arrowpos=0.8,arrowlength=5,arrowwidth=2,arrowinset=0.2](-52,72)(-98,24)
    \Line[arrow,arrowpos=0.2,arrowlength=5,arrowwidth=2,arrowinset=0.2](6,24)(-40,72)
    \Line[arrow,arrowpos=0.5,arrowlength=5,arrowwidth=2,arrowinset=0.2](-15,45)(-77,45)
    \GOval(-28,60)(7,7)(0){0.882}
    \GOval(-64,60)(7,7)(0){0.882}
    \Vertex(-15,45){4}
    \Vertex(-77,45){4}
    \Text(30,47)[lb]{\scalebox{1}{$-$}}
    \Line[arrow,arrowpos=0.5,arrowlength=5,arrowwidth=2,arrowinset=0.2](142,72)(86,72)
    \Line[arrow,arrowpos=0.5,arrowlength=5,arrowwidth=2,arrowinset=0.2](86,72)(62,24)
    \Line[arrow,arrowpos=0.5,arrowlength=5,arrowwidth=2,arrowinset=0.2](166,24)(142,72)
    \Vertex(86,72){4}
    \Vertex(142,72){4}
    \Vertex(62,24){2}
    \Vertex(166,24){2}
    \Text(46,8)[lb]{\scalebox{0.8}{$\bar\psi(x_2)$}}
    \Text(170,8)[lb]{\scalebox{0.8}{$\psi(x_3)$}}
     \Text(165,28)[lb]{\scalebox{1}{$=\frac{C_{s^2\bar\psi\psi}^{(1/N)}}{C_{s^2}}\frac{(1+\delta V_{s^2ss}+2A_s+\delta w)
    x_{23}^\mu \gamma_\mu}{\mu^{2\gamma_s-\gamma_{s^2}}\,|x_{23}|^{d-2+2\gamma_s-\gamma_{s^2}}}\Bigg|_{\frac{1}{N^2}}$}}
  \end{picture}
\end{center}
Here the extra term $\delta w$ originates as follows.
Integrating over the unique vertices of the $s^2 s s$ conformal triangle in the first diagram
on the l.h.s. of the last equation we obtain
\begin{center}
  \begin{picture}(120,72) (59,15)
    \SetWidth{1.0}
    \SetColor{Black}
    \Line[arrow,arrowpos=0.5,arrowlength=5,arrowwidth=2,arrowinset=0.2](152,72)(96,72)
    \Line[arrow,arrowpos=0.5,arrowlength=5,arrowwidth=2,arrowinset=0.2](96,72)(72,24)
    \Line[arrow,arrowpos=0.5,arrowlength=5,arrowwidth=2,arrowinset=0.2](176,24)(152,72)
    \Vertex(96,72){4}
    \Vertex(152,72){4}
    \Vertex(72,24){2}
    \Vertex(176,24){2}
    \Text(95,80)[lb]{\scalebox{0.8}{$d-1+2\gamma_s-\gamma_{s^2}$}}
  \end{picture}
\end{center}
Further taking the integrals over the last two vertices using the propagator merging relations, and
expanding the result in $1/N$ we arrive at
\begin{align}
\frac{C_{s^2\bar\psi\psi}^{(1/N)}}{C_{s^2}}\,(1+\delta w)&=\frac{C_\psi^3}{N}\,
(-\pi^d)A\left(1-\gamma_s+\frac{\gamma_{s^2}}{2}\right)V\left(\frac{d-1}{2},
\frac{d-1-\gamma_{s^2}}{2}+\gamma_s\right)\notag\\
& A\left(\frac{d-\gamma_{s^2}}{2}-1+\gamma_s\right)
V\left(\frac{d-1}{2},\frac{3+\gamma_{s^2}}{2}-\gamma_s\right)\,,
\end{align}
where we defined
\begin{equation}
\label{delta w}
\delta w = \frac{1}{2}\frac{d-4}{d-2}\, (2 \gamma _s-\gamma _{s^2})\,.
\end{equation}
Finally, we obtain
\begin{equation}
\label{delta V s2 ss result}
\delta V_{s^2ss} =  W_3+W_4-\delta w-2A_s
\end{equation}
and therefore due to (\ref{delta Zs2ss in terms of delta Vs2ss})
\begin{equation}
\label{delta Z s2 ss result}
\boxed{
\delta Z_{s^2ss} =  W_3+W_4-\delta w-2A_s-\delta z
}
\end{equation}
where $\delta z$ was calculated in (\ref{delta z expression}).

\section{$\langle s^2 s s\rangle$}
\label{sec:anomalous dim of s2}

In this section we are going to calculate
the $\langle s^2 s s\rangle$ and $\langle s^2 s^2\rangle$
correlation functions
at the next-to-leading order in the $1/N$ expansion.
In particular, we will obtain expression for the $1/N$ correction $A_{s^2}$
to the propagator amplitude of the composite operator $s^2$.
Besides deriving these new results, we will also reproduce the known expression for 
the anomalous dimension of the composite operator $s^2$.

Up to the next-to-leading order in $1/N$, the effective $\langle s^2 s s\rangle$ 
three-point function is determined by the the following diagrams
\begin{equation}
  \begin{picture}(494,129) (8,10)
    \SetWidth{1.0}
    \SetColor{Black}
    \Line[](58,67)(30,24)
    \Line[](83,24)(58,67)
    \GOval(43,45)(7,7)(0){0.882}
    \GOval(71,45)(7,7)(0){0.882}
    \CBox(55,65)(62,72){Black}{Black}
    \Vertex(30,25){2}
    \Vertex(83,25){2}
    \Text(55,77)[lb]{\scalebox{0.8}{$x_1$}}
    \Text(25,13)[lb]{\scalebox{0.8}{$x_2$}}
    \Text(82,13)[lb]{\scalebox{0.8}{$x_3$}}
    \Text(50,-5)[lb]{\scalebox{1}{$\textbf{(a)}$}}
    \Text(105,59)[lb]{\scalebox{0.8}{$+$}}
    \CBox(163,124)(170,131){Black}{Black}
    \Text(163,137)[lb]{\scalebox{0.8}{$x_1$}}
    \Line[arrow,arrowpos=0.4,arrowlength=5,arrowwidth=2,arrowinset=0.2](126,24)(166,95)
    \Line[arrow,arrowpos=0.6,arrowlength=5,arrowwidth=2,arrowinset=0.2](166,95)(208,24)
    \Line[arrow,arrowpos=0.5,arrowlength=5,arrowwidth=2,arrowinset=0.2](150,66)(184,66)
    \Line[arrow,arrowpos=0.5,arrowlength=5,arrowwidth=2,arrowinset=0.2](201,38)(134,38)
    \Vertex(134,38){4}
    \Vertex(201,38){4}
    \Vertex(150,66){4}
    \Vertex(184,66){4}
    \Vertex(126,24){2}
    \Vertex(208,24){2}
    \Line[](166,125)(166,94)
    \GOval(166,94)(8,8)(0){0.882}
    \Text(123,77)[lb]{\scalebox{0.8}{$2+\delta/2$}}
    \Text(185,77)[lb]{\scalebox{0.8}{$2+\delta/2$}}
    \Line[arrow,arrowpos=0.5,arrowlength=5,arrowwidth=2,arrowinset=0.2](275,66)(311,66)
    \Vertex(311,66){4}
    \Line[arrow,arrowpos=0.65,arrowlength=5,arrowwidth=2,arrowinset=0.2](251,24)(275,66)
    \Vertex(275,66){4}
    \Line[](335,24)(311,66)
    \Line[arrow,arrowpos=0.5,arrowlength=5,arrowwidth=2,arrowinset=0.2](310,38)(260,38)
    \Vertex(260,38){4}
    \Line[arrow,arrowpos=0.5,arrowlength=5,arrowwidth=2,arrowinset=0.2](311,66)(310,38)
    \Vertex(310,38){4}
    \Line[](310,38)(350,95)
    \Line[](275,66)(350,95)
    \Vertex(275,66){4}
    \CBox(386,124)(393,131){Black}{Black}
    \Vertex(251,24){2}
    \Vertex(335,24){2}
    \Text(386,137)[lb]{\scalebox{0.8}{$x_1$}}
    \Text(121,13)[lb]{\scalebox{0.8}{$x_2$}}
    \Text(158,-5)[lb]{\scalebox{1}{$\textbf{(b)}$}}
    \Text(243,13)[lb]{\scalebox{0.8}{$x_2$}}
    \Text(280,-5)[lb]{\scalebox{1}{$\textbf{(c)}$}}
    \Text(211,13)[lb]{\scalebox{0.8}{$x_3$}}
    \Text(337,13)[lb]{\scalebox{0.8}{$x_3$}}
    \Text(370,53)[lb]{\scalebox{0.8}{$=\langle s(x_1)^2s(x_2)s(x_3)\rangle$}}
    \Text(230,59)[lb]{\scalebox{0.8}{$+$}}
    \Line[](350,95)(390,125)
    \GOval(350,95)(8,8)(0){0.882}
    \Text(290,87)[lb]{\scalebox{0.8}{$2+\delta/2$}}
    \Text(340,67)[lb]{\scalebox{0.8}{$2+\delta/2$}}
  \end{picture}
  \label{<s^2 s s> diagrams}
\end{equation}
The diagram \textbf{(a)} of (\ref{<s^2 s s> diagrams}) represents
contributions due to the leading-order $\langle s^2 s s \rangle$
diagram where the $\langle s s\rangle$ propagators have been dressed.
Recall that we denote the dressed propagators with a solid blob,
according to the conventions introduced in section~\ref{sec:setup}.
In particular, this diagram includes the entire leading-order expression for the $\langle s^2 s s \rangle$
three-point function.

The diagrams \textbf{(b)} and \textbf{(c)} of  (\ref{<s^2 s s> diagrams}) are purely sub-leading in $1/N$.
Notice that in these diagrams we incorporated the $s^2 s s$ conformal triangle \cite{Vasiliev:1993pi}
(denoted with a solid blob, following the conventions of section~\ref{subsec:s2ss}), and
regularized the inner $s$ propagators by a small additional exponent $\delta$.
The latter will add important contributions to the finite part of the $\langle s^2 s s\rangle$
three-point function, when we expand the $s^2 s s$ conformal triangle sub-diagram
to the leading order in $1/N$.

Renormalizing the Hubbard-Stratonovich field $s$ and the composite operator $s^2$ due to
(\ref{s and s2 renormalization})
induces the counter-term diagram
\begin{center}
  \begin{picture}(94,59) (8,24)
    \SetWidth{1.0}
    \SetColor{Black}
    \Line[](58,67)(30,24)
    \Line[](83,24)(58,67)
    \CBox(55,65)(62,72){Black}{Black}
    \Vertex(30,25){2}
    \Vertex(83,25){2}
    \Text(32,47)[lb]{\scalebox{0.8}{$2$}}
    \Text(75,47)[lb]{\scalebox{0.8}{$2$}}
    \Text(55,77)[lb]{\scalebox{0.8}{$x_1$}}
    \Text(25,13)[lb]{\scalebox{0.8}{$x_2$}}
    \Text(82,13)[lb]{\scalebox{0.8}{$x_3$}}
    \Text(-30,39)[lb]{\scalebox{1}{$\frac{2\gamma_s+\gamma_{s^2}}{\delta}\times$}}
  \end{picture}
\end{center}
which will cancel the divergences.

We proceed with the calculation by integrating each term on both sides of (\ref{<s^2 s s> diagrams}) w.r.t. $x_1$.
On the r.h.s. we obtain
\begin{equation}
\label{integral of s2 ss lhs}
\int d^dx_1\,\langle s(x_1)^2s(x_2)s(x_3)\rangle = 2C_s^2\,
U(1,1,d-2)\,\left(1+h_0+\delta C_{s^2ss}\right)\,
\frac{\mu^{-2\gamma_s-\gamma_{s^2}}}{|x_{23}|^{4-d+2\gamma_s+\gamma_{s^2}}}\,,
\end{equation}
where expanding in $1/N$ we get
\begin{align}
&U\left(1+\frac{\gamma_{s^2}}{2},1+\frac{\gamma_{s^2}}{2},d-2-\gamma_{s^2}\right)
=U(1,1,d-2)\,(1+h_0)\,,\notag\\
&h_0=\gamma_{s^2}\,r\,,\label{h0}
\end{align}
where we defined
\begin{equation}
\label{r formula}
r = H_{d-3}+\pi  \cot \left(\frac{\pi  d}{2}\right)\,,
\end{equation}
while on the l.h.s. of (\ref{<s^2 s s> diagrams})
we obtain the sum of three terms due to the corresponding contributing diagrams
\textbf{(a)}, \textbf{(b)}, and \textbf{(c)}, which we represent as
\begin{align}
\int d^dx_1\,\langle s(x_1)^2s(x_2)s(x_3)\rangle &=
\frac{2C_s^2\,U(1,1,d-2)}{|x_{23}|^{4-d}}\,
\left(\left[1+h_1+2A_s-4\gamma_s\log(\mu |x_{23}|)\right]\right.\notag\\
&+\left.\left[h_2-\omega_2\,\log(\mu |x_{23}|)\right]
+\left[h_3-\omega_3\,\log(\mu |x_{23}|)\right]
\right)\,.
\end{align}
Here we grouped the terms in square brackets according to their origin from each of the three corresponding
diagrams.

Notice that each diagram contributes a finite amplitude, which we denote as $h_i$, $i=2,3$,
for the diagrams \textbf{(b)}, \textbf{(c)}, and $1+h_1+2A_s$ for the diagram \textbf{(a)}.
We decomposed the $1/N$ corrections to the latter into the contribution $2A_s$
originating from the $1/N$ correction to the amplitude of the $\langle s s\rangle$
propagators, as 
well as $h_1$ due to the $1/N$ expansion
\begin{align}
&U\left(1+\gamma_{s},1+\gamma_{s},d-2-2\gamma_{s}\right)
=U(1,1,d-2)\,(1+h_1)\,,\notag\\
&h_1=2\gamma_s\,r\label{h1}
\end{align}
of the integral over $x_1$.

Besides the finite amplitudes, each term on the l.h.s. of (\ref{<s^2 s s> diagrams})
contributes logarithmic terms due to the anomalous dimensions.
Specifically, the diagram \textbf{(a)} contributes the coefficient of $-4\gamma_s$, while we denoted
contributions of the diagrams \textbf{(b)} and \textbf{(c)} respectively as $-\omega_{2}$ and $-\omega_3$.
We will derive these expressions below in this section.

It is straightforward to calculate the integral over $x_1$
of the diagram \textbf{(b)} of  (\ref{<s^2 s s> diagrams}).
By repeatedly applying the uniqueness and propagator merging relations we obtain\footnote{The factor
of $-1$ is due to the Feynman rule for the fermion loop, and 2 is the symmetry factor.}
\begin{align}
\int d^dx_1\,\frac{\langle s(x_1)^2s(x_2)s(x_3)\rangle}{2C_s^2U(1,1,d-2)} &\supset
\frac{1}{N}\,(-2)\,C_s^2C_\psi^4\left(1+\frac{r\,\delta}{2}\right)\,\pi^d\,
A\left(\frac{d-\delta}{2}-1\right)\\
&\times V\left(\frac{3+\delta}{2},\frac{d-1}{2}\right)
A\left(1+\frac{\delta}{2}\right)V\left(\frac{d-1}{2},\frac{d-1-\delta}{2}\right)\notag\\
&\times U\left(\frac{d+\delta}{2},\frac{d-\delta}{2}-1,1\right)
U\left(1,1+\frac{\delta}{2},d-2-\frac{\delta}{2}\right)\,\frac{\mu^{-\delta}}{|x_{23}|^{4-d+\delta}}\,.\notag
\end{align}
The factor of $1+r\,\delta/2$, which we have already anticipated above, originates
from the expansion to the linear
order in $\delta$ of the integral over the position $x_1$ of the $s^2$ field, as well as
the integral over the three vertices of the $s^2 s s $ conformal triangle, while keeping only
contributions at the leading order in $1/N$.\footnote{The short-cut to reproduce $r$ is given by
\begin{align}
A\left(1{+}\frac{\delta}{4}\right)A\left(\frac{2d{-}\delta}{4}{-}1\right)
U\left(1{+}\frac{\delta}{4},1{+}\frac{\delta}{4},d{-}2{-}\frac{\delta}{2}\right)
{=}A(1)A\left(\frac{d}{2}{-}1\right)
U\left(1,1,d{-}2\right)\left(1{+}\frac{r\,\delta}{2}\right)\,,
\end{align}
where we retained only the $\delta$-dependent factors of the 
functions originating from taking the conformal integrals.}
Taking the limit $\delta\rightarrow 0$ we obtain
\begin{align}
\label{omega2}
\omega_2 &=\frac{1}{N}\,
\frac{2^d (1-\cos (\pi  d)) \Gamma \left(2-\frac{d}{2}\right) \Gamma \left(\frac{d-1}{2}\right)}{\pi ^{5/2}}\,,\\
\label{h2}
h_2&=\frac{1}{N}\,2^d\pi^{-\frac{5}{2}}\,
 \sin ^2\left(\frac{\pi  d}{2}\right) \Gamma \left(2{-}\frac{d}{2}\right)
\Gamma \left(\frac{d{-}1}{2}\right) \left(2 \,r-1\right)\,.
\end{align}

Finally, we proceed to the calculation of the non-planar
diagram, given by the diagram \textbf{(c)} of (\ref{<s^2 s s> diagrams}).
Integrating over $x_1$, the vertices of the conformal triangle, and two of the
opposite $s\bar\psi\psi$ vertices we arrive at
\begin{align}
\int d^dx_1\,\frac{\langle s(x_1)^2s(x_2)s(x_3)\rangle}{2C_s^2U(1,1,d-2)} &{\supset}
\frac{1}{N}\,({-}1)\,C_s^2C_\psi^4\left(1{+}\frac{r\,\delta}{2}\right)
\left(\pi^\frac{d}{2}A(1)V\left(\frac{d{-}1}{2},\frac{d{-}1}{2}\right)\right)^2\,\frac{v\,\mu^{-\delta}}{|x|^{4{-}d{+}\delta}}\,,
\end{align}
where we defined $v$ as an amplitude of the graph
\begin{center}
  \begin{picture}(200,118) (19,5)
    \SetWidth{1.0}
    \SetColor{Black}
    \Line[arrow,arrowpos=0.5,arrowlength=5,arrowwidth=2,arrowinset=0.2](40,64)(80,104)
    \Line[arrow,arrowpos=0.5,arrowlength=5,arrowwidth=2,arrowinset=0.2](80,104)(120,64)
    \Line[arrow,arrowpos=0.5,arrowlength=5,arrowwidth=2,arrowinset=0.2](120,64)(80,24)
    \Line[arrow,arrowpos=0.5,arrowlength=5,arrowwidth=2,arrowinset=0.2](80,24)(40,64)
    \Line[](80,104)(80,24)
    \Vertex(80,104){4}
    \Vertex(80,24){4}
    \Vertex(40,64){2}
    \Vertex(120,64){2}
    \Text(53,88)[lb]{\scalebox{0.8}{$1$}}
    \Text(104,88)[lb]{\scalebox{0.8}{$1$}}
    \Text(50,40)[lb]{\scalebox{0.8}{$1$}}
    \Text(108,40)[lb]{\scalebox{0.8}{$1$}}
    \Text(84,62)[lb]{\scalebox{0.8}{$d+\delta$}}
    \Text(26,62)[lb]{\scalebox{0.8}{$0$}}
    \Text(130,62)[lb]{\scalebox{0.8}{$x$}}
    \Text(154,56)[lb]{\scalebox{1}{$=\frac{v\,\mu^{-\delta}}{|x|^{4-d+\delta}}\,\textrm{tr}(\mathbb{I})$}}
    \Text(80,112)[lb]{\scalebox{0.8}{$x_1$}}
    \Text(80,5)[lb]{\scalebox{0.8}{$x_2$}}
  \end{picture}
\end{center}
Using the standard expression for the trace of product of four gamma matrices,
and manipulating some scalar products,
we obtain
\begin{align}
\label{expression for v}
\frac{v}{|x|^{4-d+\delta}}&= \int d^dx_{1,2}\,\frac{1}{(|x_1||x_2-x|)^2|x_{12}|^{d+\delta}}\notag\\
&-\frac{|x|^2}{2}\int d^dx_{1,2}\,\frac{1}{(|x_1||x_1-x||x_2-x||x_2|)^2|x_{12}|^{d-2+\delta}}\,.
\end{align}
The first term in (\ref{expression for v}) can be evaluated using the propagator merging relations,
while the second one can be reduced via inversion
transformation around the left-hand external point to the $\textrm{ChT}(1,1)$ integral, given by eq. (16) 
in \cite{Vasiliev:1981dg}.\footnote{
See also App. B in \cite{Goykhman:2019kcj} for a review of the derivation of this integral.} As a result we obtain
\begin{align}
\label{v rhomboid}
\frac{v\,\mu^{-\delta}}{|x|^{4{-}d{+}\delta}} &=\left[ \frac{U\left(\frac{d{+}\delta}{2},\frac{d{-}\delta}{2}{-}1,1\right)
U\left(1,1{+}\frac{\delta}{2},d{-}2{-}\frac{\delta}{2}\right)}{(\mu |x_{23}|)^{\delta}}-
\frac{1}{2}\,\textrm{ChT}(1,1)\right]\frac{1}{|x|^{4{-}d}}\,,
\end{align}
where 
\begin{equation}
\textrm{ChT}(1,1) = \pi ^d \cos \left(\frac{\pi  d}{2}\right) \Gamma (3-d) \left(\pi ^2-6 \psi ^{(1)}\left(\frac{d}{2}-1\right)\right)\,.
\end{equation}
Interestingly, only the first term in (\ref{v rhomboid}) is divergent in the $\delta\rightarrow 0$ limit.
Combining everything together we obtain
\begin{align}
\omega_3 &= \frac{1}{d-2}\,\omega_2\,,\label{omega3}\\
h_3 &=\frac{1}{N}\,\frac{2^{d-4} \Gamma \left(\frac{d-1}{2}\right)\,\sin \left(\frac{\pi  d}{2}\right)}
{\pi ^{3/2} (d-2) \Gamma \left(\frac{d}{2}\right)}\,\left(\pi ^2 (d-2)^2-16 \gamma  (d-2)
-16 \pi  (d-2) \cot \left(\frac{\pi  d}{2}\right)\right.\notag\\
&-\left.6 (d-2)^2 \psi ^{(1)}\left(\frac{d}{2}-1\right)-16 (d-2) \psi ^{(0)}(d-2)+16\right)\,.
\end{align}

Comparing the logarithmic terms on both sides of (\ref{<s^2 s s> diagrams}) we conclude
\begin{equation}
\label{gamma s2 prelim}
\gamma_{s^2} = 2\gamma_{s}+ \omega_2+\omega_3 \,.
\end{equation}
Using (\ref{gamma psi and s}), (\ref{omega2}), (\ref{omega3}), (\ref{gamma s2 prelim}) we therefore arrive at
\begin{equation}
\label{gamma s2}
\gamma_{s^2} = (2-d)\,\gamma_s\,.
\end{equation}
in agreement with the known result \cite{Gracey:1990wi}.

On the other hand, comparing the finite terms on both sides of (\ref{<s^2 s s> diagrams}) we get
\begin{equation}
\label{non-normalized delta Cs2ss}
\delta C_{s^2 s s} = h_1+2A_s+h_2+h_3-h_0\,.
\end{equation}
Consequently due to (\ref{delta Cs2ss in terms of delta V}) we derive
\begin{equation}
\label{As2 in terms of conformal triangle}
\boxed{
A_{s^2} =  h_1+h_2+h_3-h_0-\delta V_{s^2 s s}
}
\end{equation}
Using (\ref{delta V s2 ss result}) we can re-write it as
\begin{equation}
\label{As2 result}
A_{s^2} =2A_s+h\,,
\end{equation}
where we introduced
\begin{equation}
h= h_1+h_2+h_3-h_0-W_3-W_4+\delta w \,,
\end{equation}
which we can simplify as
\begin{align}
h=\frac{1}{N}\,\frac{2^{d-4} \sin \left(\frac{\pi  d}{2}\right) \Gamma \left(\frac{d-1}{2}\right) \left(8 (d-4)+\pi ^2 (d-2)-6 (d-2) \psi ^{(1)}\left(\frac{d}{2}-1\right)\right)}{\pi ^{3/2} \Gamma \left(\frac{d}{2}\right)}\,.
\end{align}

Above we have derived the next-to-leading order correction (\ref{non-normalized delta Cs2ss})
to the amplitude of the three-point function
$\langle s^2 s s\rangle$ given by (\ref{general <s^2 s s>}).
It is useful to obtain the counterpart of this expression 
for the three-point function of the normalized fields,
\begin{equation}
\label{rescaling of s and s2}
s\rightarrow  \sqrt{C_s(1+A_s)}\,s\,,\qquad
s^2\rightarrow  \sqrt{C_{s^2}(1+A_{s^2})}\,s^2\,.
\end{equation}
Using  (\ref{delta Cs2ss in terms of delta V}) we obtain
\begin{equation}
\delta  \hat C_{s^2ss} = \delta V_{s^2 s s} + \frac{A_{s^2}}{2} + A_s\,.
\end{equation}
Plugging in expression (\ref{delta V s2 ss result}) for $\delta V_{s^2 s s}$
and (\ref{As2 result}) for $A_{s^2}$ we arrive at
\begin{equation}
\label{delta hat Cs2ss result}
\boxed{
\delta  \hat C_{s^2ss} = W_3+W_4-\delta w+ \frac{h}{2}
}
\end{equation}

We can carry out consistency checks of our result (\ref{delta hat Cs2ss result})
by considering its limiting values in $d=2,4$ dimensions:
\begin{equation}
\label{d24 limits of s2ss}
\delta  \hat C_{s^2ss}|_{d=2} =-\frac{1}{2N}\,,\quad \delta  \hat C_{s^2ss}|_{d=4} = 0\,.
\end{equation}
Notice that in $d=2$ the theory is UV-free, and therefore we have $s\simeq \bar\psi ^ i\psi ^ i$.
By performing explicit contractions of the constituent fermions we obtain
\begin{align}
\langle s(x)s(0)\rangle &= 2\,N\,C_\psi^2\,\frac{1}{|x|^2}\,,\\
\langle s^2(x)s^2(0)\rangle &= 2\,N^2\,C_\psi^4\,\left(1-\frac{1}{N}\right)\,\frac{1}{|x|^4}\,,\\
\langle s^2(x_1)s(x_2)s(x_3)\rangle &=
2\,N^2\,C_\psi^4\,\left(1-\frac{1}{N}\right)\,\frac{1}{(|x_{12}||x_{13}|)^2}\,.
\end{align}
Therefore the next-to-leading in $1/N$ correction to the normalized three-point function
$\langle s^2 s s\rangle$ is given by
\begin{equation}
\delta\hat C_{s^2 s s} = -\frac{1}{2N}\,,
\end{equation}
in agreement with (\ref{d24 limits of s2ss}).

At the same time, in $d=4-\epsilon$ dimensions the UV fixed point of the GN model
 is equivalent to the IR fixed
point of the Gross-Neveu-Yukawa (GNY) model \cite{ZinnJustin:1991yn}. Such an equivalence 
implies that the CFT data of both critical theories must agree. In fact, for the GNY model,
to the leading order in the $\epsilon$-expansion we obtain $\delta\hat C_{s^2 s s} = 0$,
in agreement with (\ref{d24 limits of s2ss}).

\section{$\langle s^2\bar\psi\psi\rangle$}
\label{sec:s2psipsi}

In this section we are going to calculate the next-to-leading order correction $\delta C_{s^2\bar\psi\psi}$ to
the OPE coefficient $C_{s^2\bar\psi\psi}$.
Due to (\ref{Cs2psi psi general}) the total correction is given by
the sum of the 
correction $\delta Z_{s^2\bar\psi\psi}$ to the amplitude of the $s^2\bar\psi\psi$ conformal triangle
(\ref{expansion of Zs2psipsi def}), as well
as the amplitude corrections to the propagators of the fermion $\psi$ and
the composite operator $s^2$ attached to the conformal triangle,
\begin{equation}
C_{s^2\bar\psi\psi} = C_{s^2\bar\psi\psi}^{(1/N)} \,\left( 1 + \delta Z_{s^2\bar\psi\psi} +
2A_\psi+A_{s^2}+\delta u+  {\cal O}\left(
\frac{1}{N^2}\right)\right)\,,
\end{equation}
where $\delta u$ is given by (\ref{delta u}), and originates from integrating over the
vertices of the conformal triangle.
When calculating the OPE coefficients it is conventional to rescale
the external fields so that their propagators are normalized to unity.
Thus, rescaling $s^2$ according to (\ref{rescaling of s and s2}), and $\psi$ according to
\begin{equation}
\label{psi and s2 normalization}
\psi \rightarrow \sqrt{C_\psi\,(1+A_\psi)}\,\psi\,,
\end{equation}
we obtain the  OPE coefficient for the normalized fields
\begin{align}
\label{normalized Cs2 psi psi}
\hat C_{s^2\bar\psi\psi}&=\hat C_{s^2\bar\psi\psi}^{(1/N)} \,(1+\delta \hat C_{s^2\bar\psi\psi}) \\
&= \hat C_{s^2\bar\psi\psi}^{(1/N)} \,\left( 1 + \delta Z_{s^2\bar\psi\psi} +
A_\psi+\frac{A_{s^2}}{2}+\delta u+  {\cal O}\left(
\frac{1}{N^2}\right)\right)\,,\notag
\end{align}
where the leading order coefficient is given by \cite{Goykhman:2020ffn}
\begin{equation}
\hat C_{s^2\bar \psi \psi}^{(1/N)}
=\frac{1}{N}\,\frac{2^{d-\frac{3}{2}}\sin\left(\frac{\pi d}{2}\right)\Gamma\left(\frac{d-1}{2}\right)}
{\pi^{3/2}\,\Gamma\left(\frac{d}{2}\right)}\,,
\end{equation}
while for the $1/N$ correction we derive
\begin{equation}
\label{delta hat C s2 psi psi result}
\boxed{
\delta \hat C_{s^2\bar \psi \psi} = \delta Z_{s^2\bar\psi\psi} + A_\psi+A_s+\frac{h}{2}+\delta u
}
\end{equation}

As a consistency check for our expression (\ref{delta hat C s2 psi psi result})
let us consider its limiting value in $d=4$ dimension:
\begin{equation}
\label{d24 limits of s2psipsi}
\delta  \hat C_{s^2\bar\psi\psi}|_{d=4} =-\frac{6}{N}\,,
\end{equation}
where the UV fixed point of the GN model is critically equivalent to the IR fixed point of the GNY model.
For the latter, the $\delta  \hat C_{s^2\bar\psi\psi}$
is obtained perturbatively in $\epsilon$ in $d=4-\epsilon$
dimensions by using the fixed point value of the $s\bar\psi\psi$ coupling \cite{ZinnJustin:1991yn}
\begin{equation}
g_1^\star = 4\pi\sqrt{\frac{\epsilon}{N}}\,\left(1-\frac{3}{N}+{\cal O}(1/N^2)\right)+{\cal O}(1/N^2,\epsilon)
\end{equation}
in the leading-order $\langle s^2\bar\psi\psi\rangle$ diagram.
Since there are two $s\bar\psi\psi$ vertices in that diagram, we obtain 
$\delta \hat C_{s^2\bar\psi\psi}|_{d=4} =-6/N+{\cal O}(1/N^2,\epsilon)$,
in agreement with (\ref{d24 limits of s2psipsi}).

Finally, notice that the normalized amplitude $\hat C_{s^2\bar \psi \psi}^{(1/N)}$
vanishes in $d=2$, which can be alternatively seen as follows. In two dimensions
the GN model is asymptotically free, and $s\simeq \bar\psi\psi$. Therefore
the $\langle s^2\bar\psi\psi\rangle$ three-point function can be calculated
using the diagram
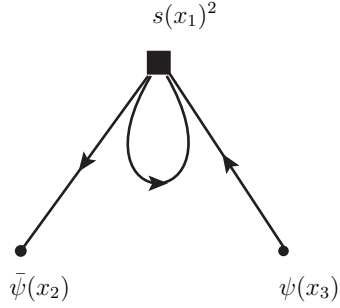
\begin{center}
  \begin{picture}(136,101) (64,-10)
    \SetWidth{1.0}
    \SetColor{Black}
    \Bezier(122,70)(153,16)(84,16)(116,70)\Line[arrow,arrowpos=0.5,arrowlength=5,arrowwidth=2,arrowinset=0.2](118.484,29.501)(118.766,29.501)
    \Line[arrow,arrowpos=0.5,arrowlength=5,arrowwidth=2,arrowinset=0.2](115,70)(67,4)
    \Line[arrow,arrowpos=0.5,arrowlength=5,arrowwidth=2,arrowinset=0.2](167,4)(123,71)
    \Vertex(67,4){2.236}
    \Vertex(166,4){2}
    \Text(117,88)[lb]{\scalebox{0.8}{$s(x_1)^2$}}
    \Text(63,-15)[lb]{\scalebox{0.8}{$\bar\psi(x_2)$}}
    \Text(165,-15)[lb]{\scalebox{0.8}{$\psi(x_3)$}}
    \CBox(115,71)(123,79){Black}{Black}
  \end{picture}
\end{center}
This diagram contains a fermionic tadpole loop, and therefore it vanishes in CFT.
In fact, $\langle s^2\bar\psi\psi\rangle|_{d=2}=0$ to all orders in $1/N$.

\section{Conformal triangle from propagator}
\label{sec:s2}

In this section we will provide an alternative derivation of 
the $1/N$ correction $\delta V_{s^2ss}$ to the amplitude of the $s^2 s s$ conformal triangle
via the $\langle s^2 s^2\rangle$ propagator. This will serve as a non-trivial consistency check for 
our result (\ref{delta Z s2 ss result}). On the other hand, the calculation presented
in this section can be viewed as a new method of deriving conformal triangles,
which we believe has not been reported before in the literature on the large $N$ vector models.

At the leading order the $\langle s^2 s^2\rangle$ two-point function is given by the diagram:
\begin{center}
  \begin{picture}(188,44) (23,-26)
    \SetWidth{1.0}
    \SetColor{Black}
    \Arc[clock](80,-68)(80,126.87,53.13)
    \Arc[](80,60)(80,-126.87,-53.13)
    \CBox(128,-8)(136,0){Black}{Black}
    \CBox(24,-8)(32,0){Black}{Black}
    \Text(24,-20)[lb]{\scalebox{0.8}{$0$}}
    \Text(128,-20)[lb]{\scalebox{0.8}{$x$}}
    \Text(176,-14)[lb]{\scalebox{1}{$=C_{s^2}\,\frac{1}{|x|^4}$}}
  \end{picture}
\end{center}
The diagrams contributing at the next-to-leading order are given by dressing
of the internal $s$ lines and the $s^2 s s$ sub-diagram of this leading order diagram. The former
is straightforward to calculate\footnote{This diagram contains the entire leading order contribution
to $\langle s^2 s^2\rangle$.}
\begin{center}
  \begin{picture}(188,44) (23,-26)
    \SetWidth{1.0}
    \SetColor{Black}
    \Arc[clock](80,-68)(80,126.87,53.13)
    \Arc[](80,60)(80,-126.87,-53.13)
    \CBox(128,-8)(136,0){Black}{Black}
    \CBox(24,-8)(32,0){Black}{Black}
    \GOval(80,-18)(10,10)(0){0.882}
    \GOval(80,12)(10,10)(0){0.882}
    \Text(24,-20)[lb]{\scalebox{0.8}{$0$}}
    \Text(128,-20)[lb]{\scalebox{0.8}{$x$}}
    \Text(176,-14)[lb]{\scalebox{1}{$=C_{s^2}\,\frac{(1+A_s)^2\,\mu^{-4\gamma_s}}{|x|^{4+4\gamma_s}}$}}
  \end{picture}
\end{center}
while the latter is given by
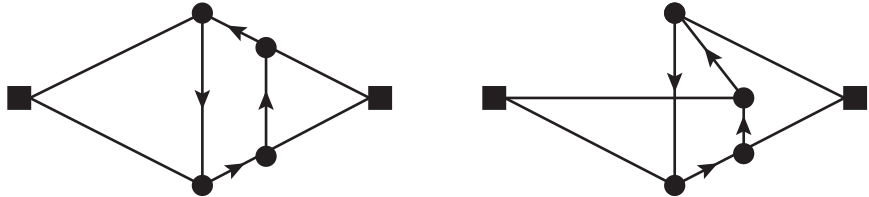
\begin{center}
  \begin{picture}(305,67) (32,-16)
    \SetWidth{1.0}
    \SetColor{Black}
    \Line[](21,18)(86,50)
    \Line[arrow,arrowpos=0.8,arrowlength=5,arrowwidth=2,arrowinset=0.2](149,18)(86,50)
    \Vertex(86,50){4}
    \Line[](21,18)(86,-15)
    \Line[arrow,arrowpos=0.2,arrowlength=5,arrowwidth=2,arrowinset=0.2](86,-15)(149,18)
    \Vertex(86,-15){4}
    \Line[arrow,arrowpos=0.5,arrowlength=5,arrowwidth=2,arrowinset=0.2](86,50)(86,-15)
    \Line[arrow,arrowpos=0.5,arrowlength=5,arrowwidth=2,arrowinset=0.2](110,-3)(110,37)
    \Vertex(110,37){4}
    \Vertex(110,-4){4}
    \CBox(13,14)(21,22){Black}{Black}
    \CBox(149,14)(157,22){Black}{Black}
    \Line[](200,18)(264,-15)
    \Vertex(264,-15){4}
    \Line[](264,50)(328,18)
    \Vertex(264,50){4}
    \Line[arrow,arrowpos=0.2,arrowlength=5,arrowwidth=2,arrowinset=0.2](264,-15)(328,18)
    \Line[arrow,arrowpos=0.40,arrowlength=5,arrowwidth=2,arrowinset=0.2](264,50)(264,-15)
    \Vertex(264,50){4}
    \Line[](200,18)(290,18)
    \Vertex(290,18){4}
    \Line[arrow,arrowpos=0.5,arrowlength=5,arrowwidth=2,arrowinset=0.2](290,-3)(290,18)
    \Vertex(290,-3){4}
    \Line[arrow,arrowpos=0.5,arrowlength=5,arrowwidth=2,arrowinset=0.2](290,18)(265,50)
    \CBox(192,14)(200,22){Black}{Black}
    \CBox(328,14)(336,22){Black}{Black}
  \end{picture}
\end{center}

Using the $s^2 s s$ conformal triangle introduced in section~\ref{subsec:s2ss} we can re-write the total of the diagrams 
contributing to $\langle s^2 s^2\rangle$ up to the next-to-leading order as
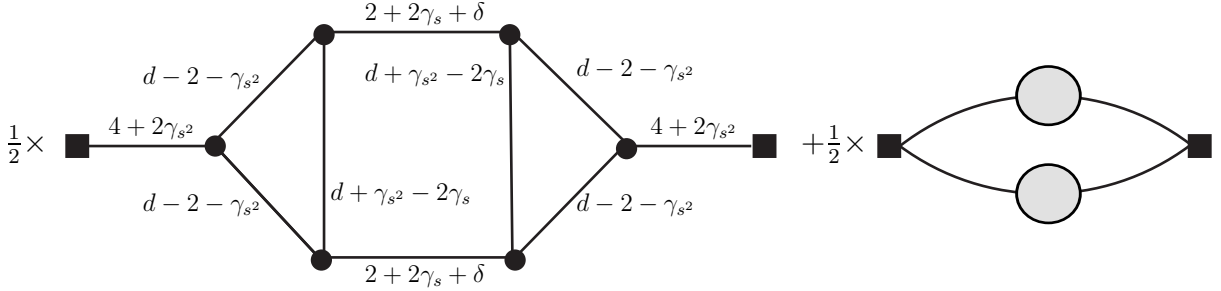
\begin{center}
  \begin{picture}(486,118) (35,-10)
    \SetWidth{1.0}
    \SetColor{Black}
    \Line[](61,57)(108,57)
    \Line[](109,57)(150,99)
    \Line[](152,100)(224,100)
    \Line[](111,57)(150,15)
    \Line[](151,15)(223,15)
    \Line[](152,100)(152,15)
    \Line[](222,100)(223,14)
    \Line[](223,100)(265,59)
    \Line[](265,57)(225,16)
    \Line[](266,57)(313,57)
    \CBox(55,53)(63,61){Black}{Black}
    \CBox(314,53)(322,61){Black}{Black}
    \Line[](111,57)(150,15)
    \Vertex(111,57){4}
    \Vertex(152,99){4}
    \Vertex(151,14){4}
    \Vertex(222,99){4}
    \Vertex(224,14){4}
    \Vertex(266,56){4}
    \Text(71,60)[lb]{\scalebox{0.8}{$4+2\gamma_{s^2}$}}
    \Text(84,79)[lb]{\scalebox{0.8}{$d-2-\gamma_{s^2}$}}
    \Text(84,31)[lb]{\scalebox{0.8}{$d-2-\gamma_{s^2}$}}
    \Text(155,35)[lb]{\scalebox{0.8}{$d+\gamma_{s^2}-2\gamma_s$}}
    \Text(168,80)[lb]{\scalebox{0.8}{$d+\gamma_{s^2}-2\gamma_s$}}
    \Text(168,103)[lb]{\scalebox{0.8}{$2+2\gamma_s+\delta$}}
    \Text(168,4)[lb]{\scalebox{0.8}{$2+2\gamma_s+\delta$}}
    \Text(248,82)[lb]{\scalebox{0.8}{$d-2-\gamma_{s^2}$}}
    \Text(248,31)[lb]{\scalebox{0.8}{$d-2-\gamma_{s^2}$}}
    \Text(276,60)[lb]{\scalebox{0.8}{$4+2\gamma_{s^2}$}}
    \Text(32,51)[lb]{\scalebox{1}{$\frac{1}{2}\times$}}
    \CBox(361,53)(369,61){Black}{Black}
    \CBox(478,53)(486,61){Black}{Black}
    \Arc[clock](424.152,-13.545)(89.545,128.018,53.033)
    \Arc[](423.5,130.444)(91.457,-126.578,-53.422)
    \GOval(425,76)(11,12)(0){0.882}
    \GOval(425,39)(11,12)(0){0.882}
    \Text(333,51)[lb]{$+\frac{1}{2}\times$}
  \end{picture}
\end{center}
Here the first diagram contains two $s^2 s s$ conformal triangles. To compensate for
this double counting we multiplied it by the factor of $1/2$.
Furthermore, the first diagram
already contains the leading order  $\langle s^2 s^2\rangle$ diagram, as well as the corrections
obtained by dressing of its internal $s$ lines. However since the first diagram
is multiplied by the factor of $1/2$, we need to add another $1/2$ of such contributions.
Finally, notice that the first diagram
is divergent. To regularize it we introduced a small shift $\delta$ to the internal $s$ propagators.

Contribution of the second diagram is given by
\begin{equation}
\label{s2s2 first contribution}
\langle s^2(x)s^2(0)\rangle \supset C_{s^2}\,\frac{1}{|x|^4}\,
\left(\frac{1}{2}+A_s-2\gamma_{s}\log(\mu |x|)\right)
\end{equation}
while contribution of the first diagram is
\begin{align}
\langle s^2(x)s^2(0)\rangle &\supset\frac{1}{2}(C_{s^2}(1+A_{s^2}) C_s (1+A_s)
Z_{s^2 s s}^{(0)}(1+\delta Z_{s^2 s s}))^2\, V(\delta)\,,
\end{align}
where $V(\delta)$ is obtained by integrating over the internal vertices.
To find the latter we first integrate over the left-most and the right-most vertices, resulting in
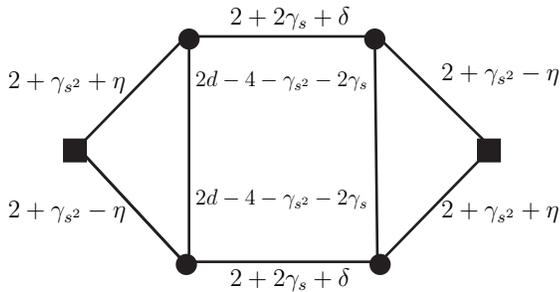
\begin{center}
  \begin{picture}(486,100) (35,6)
    \SetWidth{1.0}
    \SetColor{Black}
    \Line[](109,57)(150,99)
    \Line[](152,100)(224,100)
    \Line[](111,57)(150,15)
    \Line[](151,15)(223,15)
    \Line[](152,100)(152,15)
    \Line[](222,100)(223,14)
    \Line[](223,100)(265,59)
    \Line[](265,57)(225,16)
    \CBox(105,53)(113,61){Black}{Black}
    \CBox(261,53)(269,61){Black}{Black}
    \Line[](111,57)(150,15)
    \Vertex(152,99){4}
    \Vertex(151,14){4}
    \Vertex(222,99){4}
    \Vertex(224,14){4}
    \Text(84,79)[lb]{\scalebox{0.8}{$2+\gamma_{s^2}+\eta$}}
    \Text(84,30)[lb]{\scalebox{0.8}{$2+\gamma_{s^2}-\eta$}}
    \Text(155,35)[lb]{\scalebox{0.7}{$2d-4-\gamma_{s^2}-2\gamma_s$}}
    \Text(155,80)[lb]{\scalebox{0.7}{$2d-4-\gamma_{s^2}-2\gamma_s$}}
    \Text(168,103)[lb]{\scalebox{0.8}{$2+2\gamma_s+\delta$}}
    \Text(168,4)[lb]{\scalebox{0.8}{$2+2\gamma_s+\delta$}}
    \Text(248,82)[lb]{\scalebox{0.8}{$2+\gamma_{s^2}-\eta$}}
    \Text(248,30)[lb]{\scalebox{0.8}{$2+\gamma_{s^2}+\eta$}}
  \end{picture}
\end{center}
Here we have introduced an auxiliary parameter $\eta$, shifting exponents of some
of the lines. One can easily see that the diagram is an even function of $\eta$,\footnote{One can
see this by noticing that $\eta\rightarrow-\eta$ can be undone by swapping vertices of integration
related by mirror reflection in the horizontal axes.}
and as a result choosing $\eta = {\cal O}(\delta)$ will not affect the value of the diagram
in the limit $\delta\rightarrow 0$ \cite{Vasiliev:1981yc,Vasiliev:1981dg,Gubser:2017vgc}. We will take advantage of this fact by setting $\eta = \delta$,
which will render two of the vertices unique. Integrating over those vertices we obtain the diagram:
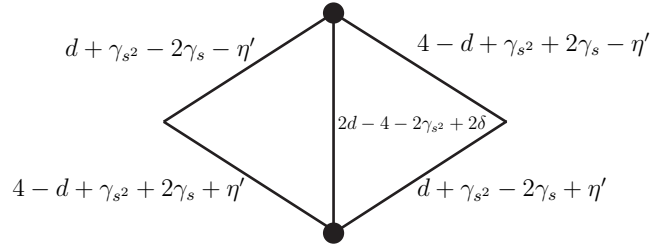
\begin{center}
  \begin{picture}(166,94) (17,-17)
    \SetWidth{1.0}
    \SetColor{Black}
    \Line[](113,70)(113,-12)
    \Line[](113,70)(49,29)
    \Line[](49,29)(113,-12)
    \Line[](113,-13)(178,29)
    \Line[](178,29)(113,70)
    \Text(-8,-1)[lb]{\scalebox{0.8}{$4-d+\gamma_{s^2}+2\gamma_s+\eta'$}}
    \Text(12,52)[lb]{\scalebox{0.8}{$d+\gamma_{s^2}-2\gamma_s-\eta'$}}
    \Text(115,25)[lb]{\scalebox{0.6}{$2d-4-2\gamma_{s^2}+2\delta$}}
    \Text(145,54)[lb]{\scalebox{0.8}{$4-d+\gamma_{s^2}+2\gamma_s-\eta'$}}
    \Text(145,-1)[lb]{\scalebox{0.8}{$d+\gamma_{s^2}-2\gamma_s+\eta'$}}
    \Vertex(113,-13){4}
    \Vertex(113,70){4}
  \end{picture}
\end{center}
Here we introduced yet another auxiliary shift $\eta'$, such that the resulting
diagram is an even function of $\eta'$.\footnote{This can be seen
by renaming the vertices of integration $x_{1,2}$ as $x_1\rightarrow x-x_2$,
$x_2\rightarrow x-x_1$. We refer the reader to \cite{Gubser:2017vgc} for the
detailed explanation of this method of calculating such diagrams.} Consequently
choosing $\eta' = \delta$ we will not change the value of the diagram
in the $\delta\rightarrow 0$ limit, while this will make the topmost vertex
unique. Completing the last two integrals we obtain for the total:
\begin{align}
V(\delta) &{=} \frac{1}{2}U\left(2{+}\gamma_{s^2},\frac{d{-}\gamma_{s^2}}{2}{-}1,\frac{d{-}\gamma_{s^2}}{2}{-}1\right)^2
U\left(1{+}\gamma_s{+}\frac{\delta}{2},d{-}2{-}\gamma_s{-}\frac{\gamma_{s^2}}{2},
1{+}\frac{\gamma_{s^2}{-}\delta}{2}\right)^2\notag\\
&\times U\left(d{-}2{-}\gamma_{s^2}{+}\delta,\frac{d{+}\gamma_{s^2}{-}\delta}{2}
{-}\gamma_s,\frac{\gamma_{s^2}{-}d{-}\delta}{2}{+}2{+}\gamma_s\right)
U\left(\frac{d}{2}{+}\delta,\frac{d}{2}{+}\delta,{-}2\delta\right)\frac{\mu^{-2\gamma_{s^2}-2\delta}}{|x|^{4+2\gamma_{s^2}+2\delta}}\,,\notag
\end{align}
where $1/2$ is the symmetry factor of the diagram. Expanding the product of the $U$ functions 
around $\delta = 0$ and $N=\infty$ we obtain
\begin{align}
V(\delta) &= v_0\left(1+\frac{\gamma_{s^2}-2\gamma_s}{\delta}
+\delta v\right)\,\frac{\mu^{-2\gamma_{s^2}-2\delta}}{|x|^{4+2\gamma_{s^2}+2\delta}}\\
&=v_0\,\left(1+\delta v+(4\gamma_s-4\gamma_{s^2})\,\log(\mu |x|)\right)\,\frac{1}{|x|^{4}}\,,\notag
\end{align}
where we subtracted divergent part using $s^2ss$ counterterm
discussed in section~\ref{sec:anomalous dim of s2}, and
\begin{align}
v_0&=\frac{C_{s^2}}{\left(C_{s^2}\,C_s\,Z_{s^2 ss}^{(0)}\right)^2}\,,\\
\delta v&=   (2 \gamma _s+\gamma_{s^2})\,\left(\pi \cot \left(\frac{\pi  d}{2}\right)
+H_{d-3}\right)-2\,\frac{d-6}{d-4}\, \gamma _{s^2}\,.
\end{align}
The corresponding contribution to the two-point function is then
\begin{equation}
\label{s2s2 second contribution}
\langle s^2(x)s^2(0)\rangle \supset C_{s^2}\,\left(\frac{1}{2}+A_{s^2}+A_s+\delta Z_{s^2 s s}+\frac{\delta v}{2}
+(2\gamma_{s}-2\gamma_{s^2})\,\log(\mu |x|)\right)\,\frac{1}{|x|^4}\,,
\end{equation}
Combining (\ref{s2s2 first contribution}), (\ref{s2s2 second contribution}) we obtain
\begin{equation}
\langle s^2(x)s^2(0)\rangle = C_{s^2}\,\left(1+A_{s^2}+2A_s+\delta Z_{s^2 s s}+\frac{\delta v}{2}\right)\,\frac{1}{|x|^{4+2\gamma_{s^2}}}\,,
\end{equation}
Using the expression (\ref{delta Z s2 ss result}) for $\delta Z_{s^2 s s}$ derived at the next-to-leading
order in $1/N$ in section~\ref{sec:s^2 bar psi psi} we obtain
\begin{align}
2A_s+\delta Z_{s^2 s s}+\frac{\delta v}{2}&= W_3+W_4-\delta w-\delta z+\frac{\delta v}{2} \notag\\
&= \frac{\gamma _{s^2}}{2}-\gamma _s+ \frac{1}{N}\,
\frac{2^d \sin \left(\frac{\pi  d}{2}\right) \Gamma \left(\frac{d+1}{2}\right)}{\pi ^{3/2} \Gamma \left(\frac{d}{2}\right)}\,.
\label{W34 consistency check}
\end{align}
Using expressions (\ref{gamma psi and s}), (\ref{gamma s2})
for the anomalous dimensions $\gamma_{s,s^2}$ we can verify explicitly
that the r.h.s. of (\ref{W34 consistency check}) vanishes,
which provides a non-trivial consistency check
for our calculation of the $W_3+W_4$ diagrams performed in section~\ref{sec:s^2 bar psi psi},
as well as for the resulting value of $\delta Z_{s^2 s s}$.

\section{Discussion}
\label{sec:discussion}

In this paper we derived new CFT data at the UV fixed point of the Gross-Neveu
model in $2<d<4$ dimensions. In particular, we further established the computational
power of the background field method, first proposed in the context of the large-$N$
vector models in \cite{Goykhman:2020ffn}, for the calculation of the finite
parts of the effective vertices, and the corresponding OPE coefficients.
To this end, we derived new expressions for the $s^2\bar\psi\psi$ and $s^2 s s$
conformal triangles, and obtained the correlation functions
$\langle s^2\bar\psi\psi\rangle$,  $\langle s^2s s\rangle$, and $\langle s^2 s^2\rangle$,
while working at the next-to-leading order in the $1/N$ expansion.

Our results are complementary to the literature on the vector models, in that
they provide the finite parts of the correlation functions, useful to obtain the
OPE coefficients. Specifically, the background field method allows one to
easily go beyond the singular parts of the correlation functions, and the associated
anomalous dimensions of the primary operators of the theory.

A natural extension of our results would be 
to apply the methods developed in the present work
to the $O(N)$ vector model
and derive the corresponding conformal triangles $s^2 s s$
and $s^2\phi\phi$, where $\phi$ is the fundamental $O(N)$
field and $s$ is the Hubbard-Stratonovich field. In particular,
the $s^2 s s$ conformal triangle can be subjected to the consistency
check by calculating the $\langle s^2 s^2 \rangle$ two-point function,
along the lines of section~\ref{sec:s2}. One can also use this data
to derive the $\langle s^2 s s \rangle$ and $\langle s^2 \phi\phi \rangle$
three-point functions and extract the corresponding OPE coefficients.

Another possible application of our results 
could be found in the study of vector models at finite temperature.
This direction of research has recently received a renewed attention due to 
the discovery of the bi-conical vector models exhibiting symmetry breaking at all
temperatures \cite{Chai:2020onq,Chai:2020zgq,Chai:2020hnu}.

\section*{Acknowledgements}

We thank Michael Smolkin for numerous discussions, comments on the draft, and the
suggestion to calculate $\langle s^2\bar\psi\psi\rangle$, and
John~Gracey for helpful correspondence. This work was partially supported by 
the
Binational Science Foundation (grant No. 2016186), the Israeli Science Foundation Center of Excellence (grant No. 2289/18) and by the Quantum Universe I-CORE program of the Israel Planning and Budgeting Committee (grant No. 1937/12).

\end{document}